\documentclass[fleqn,usenatbib]{mnras}

\usepackage{newtxtext,newtxmath}

\usepackage[T1]{fontenc}

\DeclareRobustCommand{\VAN}[3]{#2}
\let\VANthebibliography\thebibliography
\def\thebibliography{\DeclareRobustCommand{\VAN}[3]{##3}\VANthebibliography}

\newcommand{\mj}{$M_\textrm{J}$}

\newcommand{\msun}{$M_\odot$}

\newcommand{\teff}{$T_{\rm eff}$}

 \newcommand{\rev}[1]{#1}

\usepackage{graphicx}	
\usepackage{amsmath}	
\usepackage{threeparttable}
\usepackage{pdflscape}
\usepackage{booktabs}

\title[Multiplicity survey of brown dwarfs in Taurus]{An HST survey of photometry and multiplicity of brown dwarfs and planetary-mass objects in the field of Taurus}

\author[P. Liu et al.]{
Pengyu Liu,$^{1,2,3,4,5}$\thanks{E-mail: liupy6@gmail.com}
Beth A. Biller,$^{1,2}$
Cl\'emence Fontanive,$^{1,2}$
Michele Scalco,$^{6}$
Loïc Albert,$^{7,8}$
\newauthor
Luigi R. Bedin,$^{9}$
Matthew De Furio,$^{10}$\thanks{NSF Astronomy and Astrophysics Postdoctoral Fellow}
Belinda Damian,$^{11}$
Jessy Jose,$^{12}$
Zhoujian Zhang,$^{13}$
Wen-Ping Chen,$^{14,15}$
\newauthor
Sophie Dubber,$^{1,2}$
Bhavana Lalchand,$^{14,16}$
and Michael C. Liu$^{17}$
\\
$^{1}$Institute for Astronomy, University of Edinburgh, Royal Observatory, Edinburgh EH9 3HJ, UK\\
$^{2}$Centre for Exoplanet Science, University of Edinburgh, Edinburgh, UK\\
$^{3}$Leiden Observatory, Leiden University, Gorlaeus Building, Einsteinweg 55, 2333 CC Leiden, The Netherlands\\
$^{4}$Department of Astronomy, School of Science, Westlake University, Hangzhou, Zhejiang 310030, P. R. China\\
$^{5}$Institute of Natural Sciences, Westlake Institute for Advanced Study, Hangzhou, Zhejiang 310024, P. R. China\\
$^{6}$Department of Astronomy, Indiana University, Swain West, 727 E. 3rd Street, Bloomington, IN 47405, USA\\
$^{7}$D\'epartement de Physique and Observatoire du Mont-M\'egantic, Universit\'e de Montr\'eal, C.P. 6128, Succ. Centre-ville, Montr\'eal, H3C 3J7, Québec, Canada
\\
$^{8}$Observatoire du Mont-Mégantic, Université de Montréal, C.P. 6128, Succ. Centre-ville, Montréal, H3C 3J7, Québec, Canada\\
$^{9}$Istituto Nazionale di Astrofisica (INAF), Osservatorio Astronomico di Padova, Vicolo dell’Osservatorio 5, Padova I-35122, Italy\\
$^{10}$Department of Astronomy, The University of Texas at Austin, 2515 Speedway, Stop C1400, Austin, TX 78712, USA\\
$^{11}$SUPA, School of Physics \& Astronomy, University of St Andrews, North Haugh, St Andrews, KY16 9SS, UK\\
$^{12}$Department of Physics, Indian Institute of Science Education and Research Tirupati, Yerpedu, Tirupati - 517619, Andhra Pradesh, India\\
$^{13}$Department of Physics \& Astronomy, University of Rochester, Rochester, NY 14627, USA\\
$^{14}$Institute of Astronomy, National Central University, 300 Zhongda Road, Zhongli 32001 Taoyuan, Taiwan\\
$^{15}$Department of Physics, National Central University, 300 Zhongda Road, Zhongli 32001 Taoyuan, Taiwan\\
$^{16}$Yunnan Observatories, Chinese Academy of Sciences, Kuming 650216, People's Republic of China\\
$^{17}$Institute for Astronomy, University of Hawai'i, 2680 Woodlawn Drive, Honolulu HI 96822, USA
}

\date{Accepted 2026 October 04. Received 2026 September 18; in original form 2026 June 23}

\pubyear{\the\year{}}

\begin{document}
\label{firstpage}
\pagerange{\pageref{firstpage}--\pageref{lastpage}}
\maketitle

\begin{abstract}
We present a photometric and multiplicity survey of 26 young objects with masses between 6\,\mj{} and 37\,\mj{} and spectral type later than M8 in the Taurus star-forming region with the Hubble Space Telescope.
We search for companion candidates using two methods: a dedicated photometric technique based on the combination of colours from multi-band photometry and point spread function fitting. 
No companion candidates are detected beyond 30\,au.
With an updated membership and age identification, we determine luminosity, temperature, and mass for the 26 young objects using evolutionary models.
We demonstrate that our HST observations are sensitive to companions down to masses of 1\,\mj{} and even lower for some targets by artificial star injection tests.
After survey completeness correction, we derive a binary frequency $<$6.5\% for this sample in the 68\% confidence interval for companions with mass ratios between 0.1 and 1 and separations between 30 and 7,000\,au.
We find no significant difference in binary frequency among planetary-mass objects in Taurus, Upper Scorpius, and the field.
These findings support that the binary frequency of planetary-mass objects is settled at young ages, probably younger than 1--10\,Myr, suggesting that the formation of wide-orbit planetary-mass binaries is inefficient and the disruption of them likely occurs at a very early stage in a low density environment.

\end{abstract}




\section{Introduction}
The formation of brown dwarf binaries and planetary-mass binaries remains mysterious, although several formation channels have been proposed, including direct collapse and turbulent formation of dense cores resembling the formation of stellar binaries \citep{Larson1972, Goodwin2004, Andre2019}, gravitational instability in discs \citep{Adams1989, Bonnell1994, Kratter2016}, and dynamical capture \citep{Ostriker1994, Moeckel2007, Bate2012} and triple-system disruption \citep{Reipurth2015}.
Each mechanism leaves imprints on the multiplicity fraction, separation distribution, mass-ratio distribution, and eccentricity of the resulting population.
Thus, multiplicity is a key diagnostic of formation processes.

The fraction of objects found in binary or multiple systems in the field decreases with the mass of the primary: about half of solar-type FGK stars are in multiple systems \citep{Raghavan2010}, compared to 20--30\% of M dwarfs \citep{Janson2012, Ward-Duong2015}, and less than 20\% for brown dwarfs and planetary-mass objects \citep{Bouy2003,Fontanive2018, Fontanive2023}. 
Understanding the origin of this decline is critical for distinguishing whether stars, brown dwarfs, and free-floating planetary-mass objects form through common or distinct mechanisms.
Moreover, because this trend is established primarily for field populations (age $>1$ Gyr), it remains unclear whether it is primordial or the result of dynamical evolution. Surveying brown dwarf and planetary-mass binaries across different evolutionary stages is therefore crucial for constraining their formation pathways.

Although the first multiplicity surveys of very low-mass stars and brown dwarfs date back to the 1990s \citep{Fischer1992}, the population of known binaries in this regime has remained small.
Nevertheless, observations reveal that very low-mass binaries differ from stellar binaries in both orbital separation and mass ratio distributions \citep{Duchene2013, Offner2023}.
\cite{Burgasser2007} found that for systems with primary masses below 0.1\,\msun{} in the field, their mass ratios strongly peak near unity, even after correcting for observational biases.
The orbital separation distribution of very low-mass binaries is narrower than higher mass systems: the separation of the majority of very low-mass binaries is found within 10\,au via spectroscopic observations (spectroscopic binaries) and high angular resolution observations (visual binaries) \citep{Reid2002, Basri2006, Burgasser2012, Maxted2005, Allen2007}.
This indicates that brown dwarf and planetary-mass binaries are more likely to be found in equal-mass binaries at close separations, suggesting formation channels that favour similar-mass pairing with the greater dynamical stability of such configurations.

However, a handful of cases of wide binaries with separation larger than 10\,au have been discovered by direct imaging, such as SDSS J1416+13AB \citep[75\,\mj{}, L7 + 30\,\mj{}, T7.5, 45--135\,au, >1\,Gyr;][]{Burningham2010}, 2MASS J04414489+2301513 \citep[20\,\mj{}, M8.5 + 5--10\,\mj{}, M9.5--L0, 15\,au, $\sim$1\,Myr;][]{Todorov2010}, and Oph 98 AB \citep[15\,\mj{}, M9--L1 + 8\,\mj{}, L2--L6, 200\,au, 1--7\,Myr;][]{Fontanive2020}.
These systems raise more questions about the formation of very low-mass binaries in extreme cases.
Wide, very low-mass binaries have low binding energy and are subject to dynamical disruption after formation. 
Comparing very low-mass binary populations of different ages can provide clues on how dynamical evolution shapes the binary fraction.
\cite{Fontanive2018,Fontanive2023} constrained the binary fraction of late-T and Y-type brown dwarfs in the field with age $>$ 1\,Gyr to be smaller than 4.9\%.
\cite{DeFurio2022} found that the Orion Nebula Cluster \citep[a mean age of $\sim$2.2\,Myr;][]{Reggiani2011} has an excess of very low-mass binaries compared to the field, suggesting that dynamical scattering might disrupt very low-mass binaries after their formation and wide brown dwarf binaries are more common results of the formation process than we previously assumed.

The stellar environment, such as stellar density, may also have impacts on the observed multiplicity \citep[e.g.][]{Offner2023}.
Comparing multiplicity in various regions can examine the environment’s impacts on binary formation and evolution.
Simulations have found that high stellar density can disrupt binaries \citep{Ivanova2005}.
Young, sparse regions therefore offer a window into primordial multiplicity before dynamical evolution erases the initial imprint of formation. 
As one of the youngest star-forming regions in the solar neighbourhood, Taurus is a low-density environment consisting of sub-regions with an age of 1--10\,Myr \citep{Kenyon2008,Luhman2023} as well as older moving groups with ages of 10--60\,Myr \citep{Kraus2017, ZhangZ2018, LiuJ2021}.
A number of young planetary-mass objects have been discovered in Taurus in recent years \citep{Esplin2017,Luhman2017,Best2017,ZhangZ2018, Esplin2019}, providing a unique opportunity to probe the multiplicity of planetary-mass objects in a young low-density environment.

In this work, we present a photometric and multiplicity survey of brown dwarfs and planetary-mass objects in the field of Taurus. We describe sample selection, age, and membership refinement in Sect.~\ref{sec:sample_selection} and observations in Sect.~\ref{sec:taurus_obs}. We detail our photometry analysis and mass estimation of the primary targets in Sect.~\ref{sec:taurus_phot}. The multiplicity analysis, including companion search, survey completeness and companion frequency is presented in Sect.~\ref{sec:taurus_mul}.
We discuss the results in Sect.~\ref{sec:taurus_dis} and conclude this work in Sect.~\ref{sec:taurus_con}.

\section{Sample Selection, Membership, and Age Refinement}
\label{sec:sample_selection}

\subsection{Sample selection}
\begin{figure}
 \centering
	\includegraphics[width=1\columnwidth]{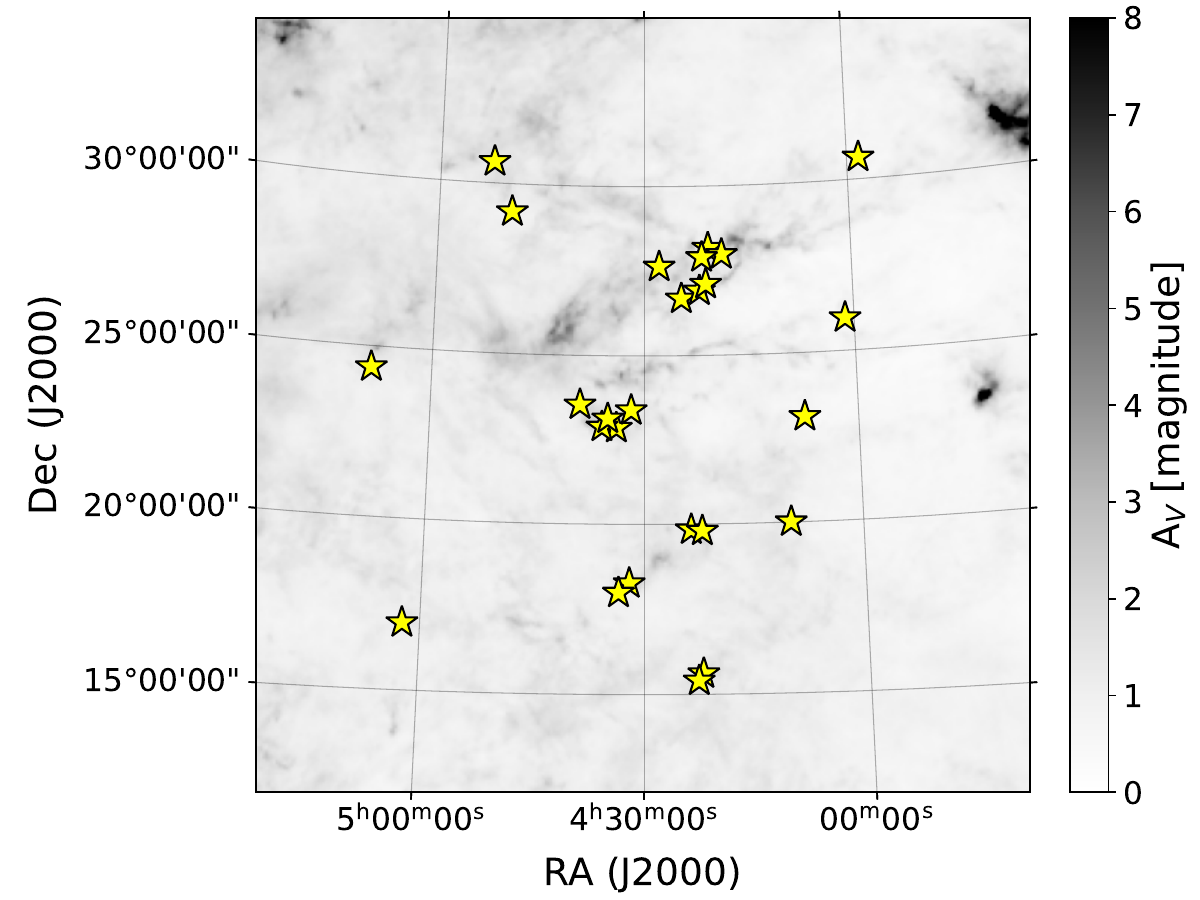}
    \caption[Taurus extinction map]{The extinction map of the Taurus field from Planck dust map \citep{Planck2014}. Targets surveyed in this work are marked with star symbols.}
    \label{fig:taurus_dustmap}
\end{figure}

For the HST multiplicity survey described here, we compiled a sample of 26 young brown dwarfs and planetary-mass objects from recent searches in the field of Taurus.
The sky map of the objects is presented in Fig.~\ref{fig:taurus_dustmap}.
General information for the targets is listed in Table~\ref{tab:taurus_target_info}, including simplified names of the targets used in this work.
Until quite recently, only a small number of brown dwarfs and planetary mass objects had been detected in Taurus, as, using traditional photometric methods, these objects are difficult to distinguish from reddened background stars. 
The initial photometric wide-band surveys in this cluster were plagued by high false positive rates due to the detection of reddened background stars \citep{Luhman2004,Luhman2009}, which have very similar broad-band near-infrared colours as substellar objects and are indistinguishable from such objects without spectroscopic follow-up. Advances in search techniques in the past decade have now revealed a new population of Taurus brown dwarfs with late M spectral types (SpT) and $<$13\,\mj{} Taurus members with L spectral types, either via dedicated custom filter surveys \citep{Allers2020, Jose2020, Dubber2021, Dubber2023, Lalchand2022,Sharma2025} or multi-epoch wide-field sky surveys which identify Taurus members by confirming common space motion with the Taurus star-forming region \citep{Best2017a, ZhangZ2018, Esplin2019}. 

\begin{landscape}
\begin{table}
\caption[Taurus target information]{Target information.}
\label{tab:taurus_target_info}
\begin{tabular}{ccccccccc}
\hline\hline
Name & Short name & RA (J2000) & Dec (J2000) & SpT & Membership$^a$ & Age (Myr) & Distance (pc) & Ref.$^b$ \\
\hline
UGCS J043506.95+230750.6 & J0435+2307 & 04:35:06.95 & +23:07:50.69 & M9-L1 & --- & 1--10 & 145$\pm$15 & WBIS, Esplin19\\
2MASS J04011678+2557527 & J0401+2557 & 04:01:16.78 & +25:57:52.73 & L1 & --- & 1.5$\pm$0.5$^c$ & 145$\pm$15 & Best17\\
UGCS J042438.53+264118.5 & J0424+2641 & 04:24:38.51 & +26:41:18.95 & M9.5 & --- & 1--10 & 145$\pm$15 & WBIS, Esplin19\\
UGCS J042201.36+265512.1 & J0422+2655 & 04:22:01.36 & +26:55:12.10 & M9-L3 & --- & 1--10 & 145$\pm$15 & WBIS, Esplin17\\
2MASS J04220204+1947499 & J0422+1947 & 04:22:02.05 & +19:47:50.00 & M9 & --- & 1--40 & 145$\pm$15 & WBIS, Esplin19\\
UGCS J043555.45+225313.4 & J0435+2253 & 04:35:55.45 & +22:53:13.49 & M9-L2 & --- & 1--10 & 145$\pm$15 & WBIS, Esplin19\\
2MASS J05082480+2422518 & J0508+2422 & 05:08:24.80 & +24:22:51.89 & L2 & --- & 1--10 & 145$\pm$15 & WBIS, Best17 \\
PSO J064.6887+27.9799 & J064.68+27.97 & 04:18:45.30 & +27:58:48.47 & L0.7 & ---- & 1--10 & 145$\pm$15 & WBIS, Zhang18 \\
PSO J065.1792+28.1767 & J065.17+28.17 & 04:20:43.01 & +28:10:36.41 & L1.6 & --- & 1--10 & 145$\pm$15 & WBIS, Esplin17, Zhang18\\
2MASS J04314964+2322524 & J0431+2322 & 04:31:49.65 & +23:22:52.42 & M9-L2 & --- & 1--10 & 145$\pm$15 & WBIS, Esplin19\\
UGCS J043328.47+175833.0 & J0433+1758 & 04:33:28.48 & +17:58:33.05 & M9-L1 & L1551(93.0\%) & 7.0$\pm$2.0 & 145.6$\pm$3.8 & WBIS, Esplin19\\
PSO J061.8714+23.1104 & J061.87+23.11 & 04:07:29.12 & +23:06:38.18 & M9.9 & --- & 1--10 & 145$\pm$15 & Zhang18 \\
2MASS J04095154+2000428 & J0409+2000 & 04:09:51.55 & +20:00:42.70 & M9.1 & --- & 1--10 & 126.6 $\pm$ 13.5 $^e$ & Zhang18\\
PSO J075.5825+16.9498 & J075.58+16.94 & 05:02:19.82 & +16:56:59.32 & M9.5 & --- & 10--30$^d$ & 145$\pm$15 & Zhang18\\
PSO J059.5714+30.6327 & J059.57+30.63 & 03:58:17.16 & +30:37:58.03 & M9.7 & --- & 10--30$^d$ & 145$\pm$15 & Zhang18\\
2MASS J04274951+2738155 & J0427+2738 & 04:27:49.52 & +27:38:15.57 & M9.5 & --- & 1--10 & 145$\pm$15 & Esplin17 \\
2MASS J04320157+1815229 & J0432+1815 & 04:32:01.57 & +18:15:22.96 & M9.5 & --- & 1--40 & 145$\pm$15 & WBIS, Esplin17\\
UGCS J043354.07+225119.1 & J0433+2251 & 04:33:54.07 & +22:51:19.20 & M9-L2 & --- & 1--10 & 145$\pm$15 & WBIS, Esplin17\\
UGCS J045210.35+303734.3 & J0452+3037 & 04:52:10.35 & +30:37:34.30 & M9.5 & L1536(90.2\%) & 3.3$\pm$0.9 & 159.2$\pm$4.2 & WBIS, Esplin17\\
2MASS J04210636+2706545 & J0421+2706 & 04:21:06.36 & +27:06:54.57 & M9.5 & --- & 1--10 & 145$\pm$15 & WBIS, Esplin19\\
2MASS J04213847+2754146 & J0421+2754 & 04:21:38.48 & +27:54:14.62 & M9.5 & --- & 1--10 & 145$\pm$15 & WBIS, Esplin19\\
WBIS2241+1523 & J0422+1523 & 04:22:41.77 & +15:23:53.67 & L0.0 & --- & 1--40 & 145$\pm$15 & WBIS\\
WBIS3902+2332 & J0439+2332 & 04:39:02.38 & +23:32:12.74 & M9.0 & L1529(96.4\%) & 3.4$\pm$0.5 & 129.4$\pm$2.7 & WBIS\\
(2MASS J04390237+2332133) & & & &  & & & \\
WBIS2204+1536 & J0422+1536 & 04:22:04.48 & +15:36:20.56 & M8.0 & --- & 1--40 & 145$\pm$15 & WBIS\\
PSO J065.8871+19.8386 & J065.88+19.83 & 04:23:32.91 & +19:50:19.26 & L0 & --- & <1$^d$ & 151.8 $\pm$ 6 $^e$ & Zhang18\\
2MASS J04492210+2911124 & J0449+2911 & 04:49:22.11 & +29:11:12.26 & M8.5 & --- & 1--10 & 145$\pm$15 & Esplin17\\
\hline
\end{tabular}
\begin{tablenotes}
\item Notes: (a).This work. Only targets with a probability $>$ 80\% are assigned to sub-regions in Taurus. 
(b).References of sources: Best17: \cite{Best2017a}, Esplin17: \cite{Esplin2017}, Zhang18: \cite{ZhangZ2018}, and Esplin19: \cite{Esplin2019}.
(c).Best17.
(d).Zhang18.
(e).Gaia DR3
\end{tablenotes}
\end{table}
\end{landscape}

One efficient manner of searching for these important young objects is to identify them via spectral features, such as the 1.45 $\mu$m H$_2$O absorption band seen in the spectra of M-L-T-Y type objects. Combining 1.45 $\mu$m imaging with $J$ and $H$ photometry robustly estimates spectral types independent of reddening \citep{Allers2020}. A 27 square degree survey of several young star-forming regions was conducted using a custom 1.45 $\mu$m filter (henceforth the $W$ filter) with the Canada France Hawaii Telescope (CFHT) in order to detect the lowest mass members of these clusters using this technique \citep{Allers2020, Jose2020, Dubber2021, Dubber2023, Lalchand2022, Sharma2025, Esplin2019}.  This survey, henceforth the Water Band Infrared Survey (WBIS), obtained $J$ \rev{(central wavelength = 1.252\,$\mu$m and bandwidth = 158\,nm)}, $H$ \rev{(central wavelength = 1.631\,$\mu$m and bandwidth = 289\,nm)}, and $W$ \rev{(central wavelength = 1.453\,$\mu$m and bandwidth = 88.4\,nm)} imaging in the Taurus field with CFHT Wide-field InfraRed Camera (WIRCam) \citep{Puget2004} (CFHT programs 15BC11, 16AC13, 16AF16, 16AT04, 16BC17, 16BF22, 17AS05, 17BC16, 17BF11,
17BT04), followed by spectroscopic confirmation of likely late M and L type members of Taurus using the NASA Infrared Telescope Facility (IRTF)/SpeX \citep{Rayner2003} or Gemini Near-Infrared Spectrograph \citep[GNIRS;][]{Elias2006}. Biller et al. in prep present the full WBIS survey design and overall results. We include in our sample 15 late M and L type members of Taurus that were identified in WBIS data by \citet{Esplin2017,Esplin2019}; we report three new WBIS Taurus brown dwarf discoveries in this paper.  The rest of the sample is selected from recent multi-epoch wide-field surveys \citep{Best2017a,Luhman2017,Esplin2017,ZhangZ2018}.  As of the HST measurement epoch, this was as near complete a census of Taurus brown dwarfs as we were able to assemble. 

\subsection{New Brown Dwarf Discoveries from the $W$-band Infrared Survey}

The $J$, $H$, and $W$ band imaging obtained with WBIS enabled a rough estimate of object spectral type for M and L type objects, based on the strength of the 1.45 $\mu$m water absorption feature in any given object.  Following \citet{Allers2020}, in combination with contemporaneous $J$ and $H$ band photometry, one may construct a reddening-insensitive index, $Q$, to probe the depth of this absorption feature while taking into account reddening due to dust extinction in star-forming regions.  \citet{Allers2020} define $Q$ as:
 
\begin{equation}
    Q = (J-W) + \frac{E(J-W)}{E(H-W)}(H-W) = (J-W) + e(H-W),
\label{eq_Q}
\end{equation}

where the constant $e$ is given by ratio of $(J-W)$ and $(H-W)$ colour excesses.  The value of the constant, $e$, depends on the adopted reddening law and spectral type of the object observed.  The typical contaminant found in Taurus are reddened background stars with early M spectral types, and thus, \cite{Allers2020} construct $Q$ in such a way that such a reddened M0 will have a $Q$ value of 0, whereas later spectral type objects with deepening water absorption features will have progressively more negative values of $Q$, adopting a value of $e$ = 1.85, the appropriate value for an M0 star reddened by A$_V$ = 10, adopting R$_V$ = 3.1 from \cite{Fitzpatrick1999}. $Q$ provides a rough proxy for spectral type, with progressively more negative $Q$ values corresponding to later spectral types.  As noted above, $Q$ is anchored to a value of 0 for M0 stars; objects with SpT later than M6 will have $Q\leq$-0.6.

While a $Q$-index value $<$-0.6 is a strong indicator of water absorption in a given candidate, it does not indicate youth or Taurus membership --- foreground and background objects with SpT later than M6 will also be detected using this $Q$-index-based selection criterion. Objects in the WBIS fields with $Q\leq$-0.6 were followed up spectroscopically with either IRTF/SpeX or Gemini GNIRS to confirm spectral type and youth (Biller et in prep); most of the WBIS-identified new young members of Taurus from these spectroscopic observations were independently discovered and published in \citet{Esplin2019}.  Here, we report the discovery of three additional young brown dwarfs co-located with Taurus.  The spectroscopic observations of these three objects are detailed in Table~\ref{tab_specobservations}.  Additional follow-up spectroscopy for WBIS will be described in Biller et al. in prep.  

WBIS~$2241+1523$ was observed at Gemini North using GNIRS \citep{Elias2006} in its cross-dispersed mode with the 32\,$\ell$\,mm$^{-1}$
grism and the short camera configuration to obtain near infrared (NIR) spectra between 0.9 and 2.5\,\micron{} split into 5 diffracting orders at a resolving power of 1700. We used a 1.0\arcsec{} slit and an ABBA dither pattern between exposures, with 8 exposures of 277 seconds, and 4 short exposures for the bright flux calibrators, and with an A0V observed after our science target at approximately the same air mass as our science target.  Spectra were extracted and flux-calibrated using our own scripts written in the IDL language described in \cite{Delorme2008}. 

WBIS~$3902+2332$ and WBIS~$2204+1536$ were observed at the Infrared Telescope Facility (IRTF) using the SpeX instrument \citep{Rayner2003} in PRISM mode ($R\sim200$) using the 0.8$\times$15\arcsec{} slit, as part of program 2018B053 (PI Biller).  Observations were taken with an ABBA dither pattern, with the A2V star HD283593 observed directly before or after each of these targets, and flat-field and argon lamps were taken at the beginning and end of each night.  Spectra were extracted and flux-calibrated using the SPEXTOOL package (\citealt{Vacca2003, Cushing2004}).

We estimated spectral types and extinction for these three brown dwarfs via comparison to the young templates of \cite{Luhman2017} that span the range of spectral types from M0.5 to M9.5 (in steps of 0.5 sub-types) and included four templates in the L dwarf regime (L0, L2, L4, L7). We also compared our spectra with those of 14 high-gravity field templates from \cite{Burgasser2014b} spanning the range from M4 to M9 and from L0 to L8. We fit all templates to each observed spectrum, leaving two parameters free: the observation-to-template relative scale and the extinction, $A_V$. 
For each candidate, we used the \texttt{scipy.optimize.least\_squares} algorithm to find the best scale and $A_V$ parameters for each template, and then plotted maps of the resulting reduced $\chi^2$ for all combinations of template and $A_V$ considered.  The template and value of $A_V$ which minimise the reduced $\chi^2$ are adopted as the best fitting spectral type, age range, and $A_V$, while the reduced $\chi^2$ lets us determine whether there are multiple well-fitting solutions as well as the uncertainty range on our best template and $A_V$.

\begin{figure}
 \centering
	\includegraphics[width=1\columnwidth]{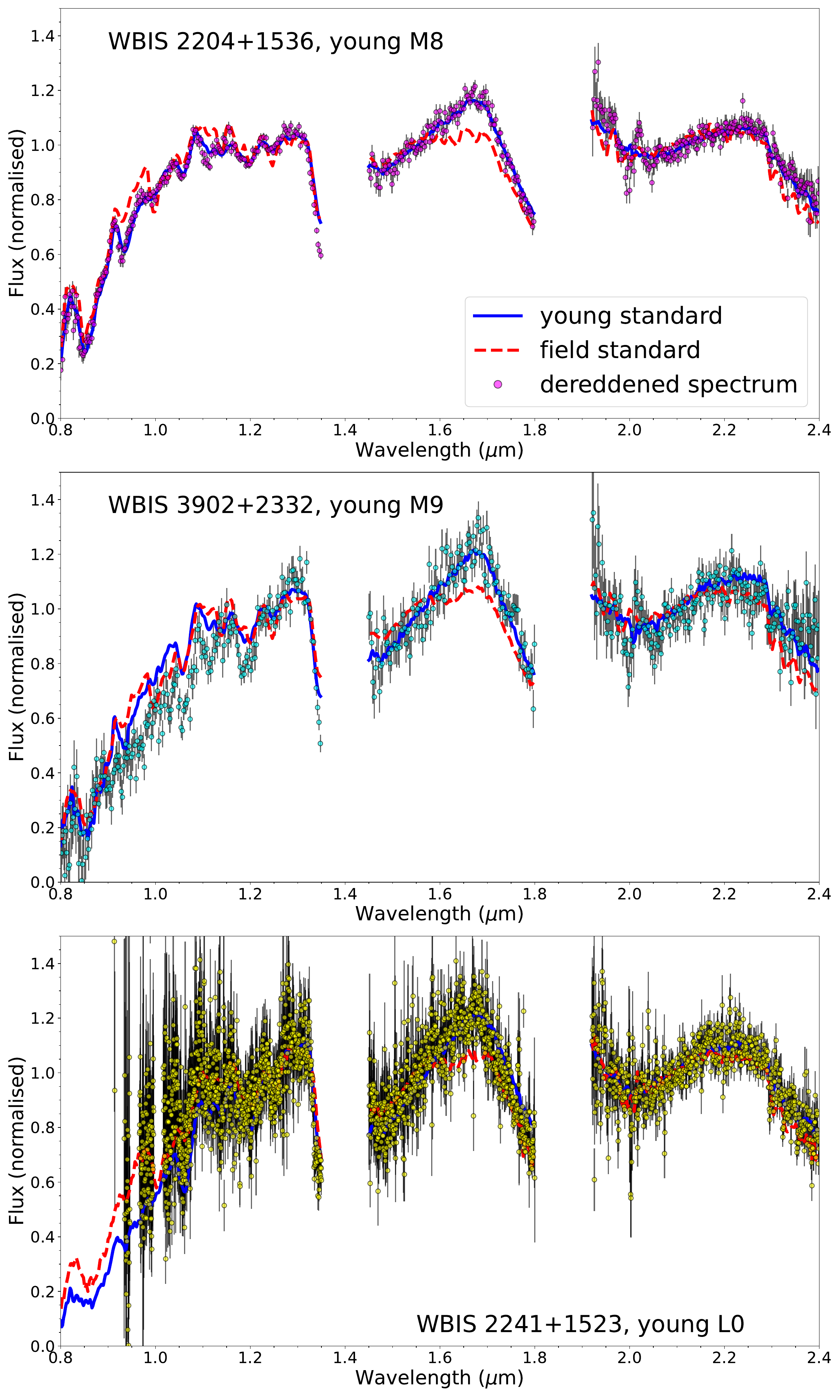}
    \caption[New brown dwarfs]{Dereddened spectra of newly discovered young brown dwarfs co-located with the Taurus star-forming region.  Each spectral band ($J$, $H$, and $K$) has been separately normalised, and noisy regions between bands have not been plotted here.  We overplot the best-fit young spectral template (blue lines, from \cite{Luhman2017}), as well as an old field template at the same spectral type (red dashed lines, from \cite{Burgasser2014}).  The young templates provide notably better fits in $H$ band in particular, capturing the peaky $H$ band feature which is a clear indicator of youth for late M and L-type objects.}
    \label{fig:bdspectra}
\end{figure}

All three young brown dwarfs presented here were best fit with young templates and moderate values of $A_V$.  The spectra, after de-reddening by the best-fit values of $A_V$, are presented in Fig.~\ref{fig:bdspectra}.  Spectra for all three objects display clear indications of youth, most notably, a `peaky' $H$ band morphology.  While the spectral fits rule out old ages for all three objects, there is significant degeneracy between extinction and spectral type for these objects.  As a result, the uncertainty of the spectral types presented for these objects is on the order of $\pm$1.5 spectral types.  All three of these new young objects have SpT later than M6, but the exact spectral type measured for each is less precise.  

The best-fit parameters from the template fits, as well as the measured $Q$-parameter, $z-J$ colours \citep{Tonry2012}, \textit{WISE}1 -- \textit{WISE}2 colours \citep{Wright2010}, are shown in Table~\ref{tab_newbd}. Brown dwarfs and planetary mass objects possess extremely red colours.  At the stellar and substellar boundary of M6 in Taurus, $z-J\approx 2.0$\,\rev{mag}, but extinction by interstellar medium dust can also produce an artificially reddened $z-J$ colour. Wide-field Infrared Survey Explorer (WISE) colours can also be used to discriminate between likely members and contaminants. \citet{Kirkpatrick2011} find $W1-W2$ $>$ 0.1\,\rev{mag} for objects later than M5 and $W1-W2$ $>$ 0.3\,\rev{mag} for objects later than L5.  The spectral fits which show clear indications of youth, measured $Q <$ $-$1\,\rev{mag} as well as the red $z-J$ and $W1-W2$ colours paint a consistent picture for all three objects as bonafide, young substellar objects.  They are co-located on the sky with the Taurus star-forming region, but, while their youth is confirmed, this analysis does not necessarily confirm Taurus membership.

\begin{table*}

\caption{Table of spectroscopic observations.\label{tab_specobservations}}

\begin{tabular}{ccccccrrc}
\hline\hline
 Target & RA & Dec & $H$ Mag. & Instrument & Date & Air Mass & Int. Time & Calibrator\\
       & (J2000)   &  (J2000)   & (Vega)      &            &      &          & (n$\times$sec) & \\
 \hline

WBIS~$2241+1523$ & 04:22:41.77 & $+$15:23:53.67 & $17.17\pm 0.08$ & GNIRS & 2018-09-11 & 1.11-1.17 & 8$\times$277 & HIP17704\\
WBIS~$3902+2332$ & 04:39:02.38 & $+$23:32:12.74 & $16.35\pm 0.04$ & SpeX & 2018-09-23 & 1.002 & 14$\times$180 & HD283593\\
WBIS~$2204+1536$ & 04:22:04.48 & $+$15:36:20.56 & $15.15\pm 0.02$ & SpeX & 2018-09-23 & 1.088 & 12$\times$120 & HD283593\\
\hline
\end{tabular}
\end{table*}

\begin{table*}
\caption{Photometry and Spectral Types for New, Young Brown Dwarf Discoveries \label{tab_newbd}}

\begin{tabular}{cccccc}
\hline\hline
 Target & $Q$ & $z-J$ (Mag) & $W1-W2$ (Mag) & SpT & $A_V$ \\
 \hline

WBIS~$2241+1523$ & -1.14$\pm$0.24 & ... & 0.7$\pm$0.1 & L0 & 3.26 \\
WBIS~$3902+2332$ & -1.12$\pm$0.13 & 3.42$\pm$0.06 & 0.69$\pm$0.06 & M9 & 0.0 \\
WBIS~$2204+1536$ & -1.36$\pm$0.05 & 2.94$\pm$0.02 & ... & M8 & 0.71 \\
\hline
\end{tabular}
\end{table*}

\subsection{Proper motion, membership and age refinement}
\label{sec:membership}
The field of Taurus is a complicated region including the Taurus star-forming region, young moving groups associated with Taurus molecular clouds and relatively older moving groups \citep{LiuJ2021, Kerr2021}. 
We classified the sub-region membership of objects that are from WBIS.
Proper motions were measured for all spectroscopically followed-up WBIS objects (18 sources in our sample). Catalogues at the following epochs were used: our CFHT/WIRCam WBIS, the Wide Field Camera (WFCAM) surveys (Galactic Clusters Survey (GCS), Galactic Plane Survey (GPS) and the UKIRT Hemisphere Survey (UHS)) \citep{Hambly2008} as well as the Pan-STARRS Surveys DR1 \citep{Chambers2016}. Around each WBIS object, sources in the catalogue were cross-matched with those of \textit{GAIA} DR2 \citep{Gaia2018} (projected at the appropriate date) within a 5-arcminute radius. The astrometry of each catalogue was fine-tuned by applying a scale and shear transformation to minimise deviations from \textit{GAIA} DR2. This yielded \rev{proper motion} (PM) measurements with typical uncertainties of 7--8 mas yr$^{-1}$, as shown in Table~\ref{tab:propermotion}.
Then, membership probabilities were calculated using \rev{Bayesian analysis tool}, Banyan $\Sigma$ \citep{Gagne2018b} with the \rev{Montreal Open Clusters and Associations} (MOCA) database \citep{Gagne2024,Gagne2026} based on the RA+DEC positions and proper motions of our sources. Probabilities were calculated for the Taurus young association at large (Prob. Taurus) and the subgroups of Taurus (Prob. Y.Assoc).
Table~\ref{tab:propermotion} presents the results. The included subgroups of Taurus are 
TAUMGLIU4 \citep[2\,Myr, 158$^{+8}_{-10}$\,pc,][]{LiuJ2021}, 
TAUMGLIU9 \citep[7--8\,Myr, 176$^{+15}_{-13}$\,pc,][]{LiuJ2021}, 
OH28 \citep[23--41\,Myr,][]{LiuJ2021}, 
OH29 \citep[29.3$\pm$\,Myr,][]{Luhman2024},
TAU \citep[1--2\,Myr,][]{Kenyon1995},
CWNU1129 \citep[15$^{+22}_{-11}$\,Myr,][]{Hunt2023}, OCSN50 \citep[20$^{+21}_{-10}$\,Myr,]{Hunt2023}, 
HSC1318 \citep[5.5$^{+2.5}_{-2.1}$\,Myr, 128\,pc,][]{Hunt2023}, 
L1551 \citep[7$\pm$2\,Myr, 145.6$\pm$3.8\,pc,][]{Kerr2021}, 
L1536 \citep[3.3$\pm$0.9\,Myr, 159.2$\pm$4.2\,pc,][]{Kerr2021}, 
L1495 \citep[3.8$\pm$0.8\,Myr, 131.2$\pm$3.3\,pc,][]{Kerr2021}, 
and L1529 \citep[3.4$\pm$0.5\,Myr, 129.4$\pm$2.7\,pc,][]{Kerr2021}.

\begin{table*}
\centering
\caption{Proper motion and probabilities of belonging to the Taurus young associations at large and a specific Taurus sub-association, as calculated using MOCA.}
\begin{tabular}{l|l|c|c|c|c}
\hline
\hline
Name & WBIS Designation  &  PM$_{\rm ra}$      &  PM$_{\rm dec}$      &  Prob. Taurus$^a$ & Prob. Y.Assoc.$^b$  \\ 
&                   & (mas yr$^{-1}$)        & (mas yr$^{-1}$)         &            (\%)    & (\%)    \\
\hline
J0435+2307 & WBIS~$3506+2307$ &  $ 9.37\pm 4.76$ & $-14.35\pm 2.13$ & 92.3       &                           TAUMGLIU4(72);CWNU1129(16)  \\
J0424+2641 & WBIS~$2438+2641$ & $ 8.87\pm13.64$ & $-11.15\pm14.75$ & 27.3       &    CWNU1129(46);HSC1318(16);L1495(14);TAU(6);L1529(6)  \\
J0422+2655 & WBIS~$2201+2655$  & $ 6.65\pm 0.69$ & $ 11.14\pm 1.35$ & 0.1       &     HSC358 \\
J0422+1947 & WBIS~$2202+1947$  & $ 2.89\pm 7.29$ & $-13.15\pm 1.80$ &  72.6       &            OCSN50(47);OH29(18);TAU(13);OH28(8);L1536(5)  \\
J0435+2253 & WBIS~$3555+2253$  & $ 2.78\pm 6.02$ & $-15.87\pm 2.88$ & 91.0       &                     TAUMGLIU4(64);L1529(13);OH28(10) \\
J0508+2422 & WBIS~$6824+2422$  & $ 4.42\pm 7.47$ & $-16.07\pm 2.85$ &  82.1       &                 TAUMGLIU9(50);TAUMGLIU6(30);L1536(5)  \\
J064.68+28.17 & WBIS~$1845+2758$  & $ 4.32\pm 2.66$ & $-26.19\pm 8.38$ & 98.3       &                          HSC1318(45);L1495(42);TAU(6)  \\
J065.17+28.17 & WBIS~$2043+2810$  & $ 6.77\pm 5.43$ & $-27.96\pm 2.57$  & 99.4       &                           HSC1318(47);L1495(42);TAU(6)  \\
J0431+2322 & WBIS~$3149+2322$ & $ 9.54\pm 4.43$ & $-14.31\pm 3.97$ & 90.9       &                          TAUMGLIU4(61);CWNU1129(25)  \\
J0433+1758 & WBIS~$3328+1758$  & $ 9.46\pm 3.85$ & $-18.35\pm 5.80$ & 93.0       &      L1551 \\
J0432+1815 & WBIS~$3201+1815$  & $17.30\pm 2.37$ & $-24.84\pm 2.48$ & 70.2       &                             TAUMGLIU4(61);CWNU1129(25) \\
J0433+2251 & WBIS~$3354+2251$  & $16.73\pm 9.88$ & $-12.46\pm 5.81$ & 85.6       &                  TAUMGLIU4(64);CWNU1129(24);L1529(5)  \\
J0452+3037 & WBIS~$5210+3037$  & $-3.78\pm 0.64$ & $-15.78\pm 2.92$ & 90.2       &      L1536 \\
J0421+2706 & WBIS~$2106+2706$  & $-3.77\pm10.67$ & $-17.27\pm 6.26$ & 89.6       &    HSC1318(29);CWNU1129(27);L1495(18);L1536(11);TAU(6)  \\
J0421+2754 & WBIS~$2138+2754$ & $15.77\pm 7.62$ & $-28.39\pm 4.24$ & 97.2       &                                    L1495(45);HSC1318(40)  \\
J0422+1523 & WBIS~$2241+1523$  & $11.08\pm 0.05$ & $ -3.41\pm 0.13$ & 0.0       &                        L1536(59);TAUMGLIU6(31);HSC358(8)  \\
J0439+2332 & WBIS~$3902+2332$  & $ 5.70\pm 1.18$ & $-23.36\pm 1.80$ & 96.4       &      L1529 \\
J0422+1536 & WBIS~$2204+1536$  & $ 4.60\pm 4.05$ & $ -5.78\pm 3.63$ & 0.7       &                              TAU(44);OH29(36);TAUMGLIU6(9) \\
\hline
\end{tabular}
\begin{tablenotes}
\item Notes: (a).Membership probabilities of young associations in Taurus at large, calculated by [Sum over all associations] / [Field + Sum over all associations].
(b).Relative membership probabilities of specific sub-associations, calculated by [One association] / [Sum over all associations].
\end{tablenotes}
\label{tab:propermotion}
\end{table*}

In general, objects with probabilities greater than 80\% are considered to have secure membership.
First, we looked into the 6 objects that have probabilities of being members of young associations in Taurus smaller than 80\%. 
All objects are clearly young from their spectra, and therefore, they are very unlikely field objects. 
J0424+2641 (WBIS~2438+2641) and J0422+2655 (WBIS2201+2655) have large uncertainties in membership classification due to very deviant astrometric data of 2013 compared to the other two epochs of 2010 and 2017.
Therefore, their probabilities are not trustworthy, and we adopted typical ages of young associations in Taurus 1--10\,Myr according to their spectra. 
J0422+1947 (WBIS~2202+1947) and J0432+1815 (WBIS~3201+1815) have probabilities of $\sim$70\% being members of young associations in Taurus, which are trustworthy as their astrometric data are not problematic. 
We adopted ages of 1--40\,Myr for them, which is a conservative range as it also includes the typical age of old moving groups in Taurus.
J0422+1523 (WBIS~2241+1523) and J0422+1536 (WBIS~2204+1536) have sub-nominal uncertainties in proper motion fitting, suggesting possibly unreliable probabilities.
We adopted ages of 1--40\,Myr for them as well.

12 objects are considered secure members of young associations in Taurus as they have probabilities over 80\%.
Among them, three have probabilities over 80\% of being members of specific sub-regions: J0433+1758 (WBIS~3328+1758), J0452+3037 (WBIS~5210+3037), and J0439+2332 (WBIS~3902+2332).
We assigned the ages and distances of the sub-regions to them.
For the other 9 objects, we adopted the typical ages of young associations in Taurus for them: 1--10\,Myr.

For the rest 8 objects, four have clear age characterisation from the literature: 2MASS J04011678+2557527 \citep[1.5 $\pm$ 0.5\,Myr,][]{Best2017}, PSO J075.5825+16.9498 and PSO J059.5714+30.6327 \citep[10--30\,Myr,][]{ZhangZ2018}, and PSO J065.8871+19.8386 \citep[$<$1\,Myr,][]{ZhangZ2018}.
For the other four, we assigned the typical Taurus young association age of 1--10\,Myr for them.
We assigned the typical distance of Taurus 145 $\pm$ 15 pc for the objects.
\rev{Two objects have Gaia~DR3 parallax measurements: 2MASSJ04095154+2000428 and PSOJ065.8871+19.8386. We used the distance calculated from the parallax.}
The adopted memberships, ages, and distances are summarised in Table~\ref{tab:taurus_target_info}.

\section{Observations}
\label{sec:taurus_obs}

The targets were observed with the Hubble Space Telescope (HST)/Wide Field Camera 3 (WFC3) in GO programme 15628 (PI Biller). They were observed in one optical filter on the UVIS detector F850LP (pivot wavelength = 0.92\,$\mu$m, width = 0.12\,nm), and two filters on the infrared (IR) detector: F127M (pivot wavelength = 1.27\,$\mu$m, width = 0.69\,$\mu$m) and F139M (pivot wavelength = 1.38\,$\mu$m, width = 0.64\,$\mu$m).
The UVIS detector has a pixel scale of 0.04\arcsec{} and a field of view (FoV) of 162$\times$162 arcsec$^2$. 
The IR detector has a pixel scale of 0.13\arcsec{} and FoV of 135$\times$135 arcsec$^2$.
The instrument settings of most targets were the same.
UGCS J043506.95+230750.6 is the only target that was observed in the full array with two dither positions because there was a candidate companion from previous Keck images. Its SAMP-SEQ and NSAMP parameters for IR observations were STEP50 $\times$ 13. 
All other targets were observed with the sub-array aperture UVIS2-2K2C-SUB (FoV = 81 $\times$ 81\,arcsec$^2$) in F850LP and SQ512SUB (FoV = 68 $\times$ 68\,arcsec$^2$) in F127M and F139M with three dither positions. Their SAMP-SEQ and NSAMP parameters were STEP25 $\times$ 14.
There are a total of 31 observations after excluding three failed observations. Four targets were observed more than once.
The valid observations of this sample are listed in Table~\ref{tab:taurus_obslog}.

\begin{table*}
\caption{HST observation summary.}
\begin{tabular}{lllll}
\hline\hline
Name & Date & F850LP Time (s) & F127M Time (s) & F139M Time (s)\\
\hline
J0435+2307 & 2019-12-03 & 478.0$\times$3 & 349.2$\times$3 & 349.2$\times$3 \\
J0435+2307 & 2019-12-17 & 478.0$\times$3 & 349.2$\times$3 & 349.2$\times$3 \\
J0401+2557 & 2019-07-30 & 270.0$\times$3 & 197.2$\times$3 & 197.2$\times$3 \\
J0424+2641 & 2019-09-02 & 270.0$\times$3 & 197.2$\times$3 & 197.2$\times$3 \\
J0422+2655 & 2019-12-14 & 270.0$\times$3 & 197.2$\times$3 & 197.2$\times$3 \\
J0422+1947 & 2019-07-26 & 267.0$\times$3 & 197.2$\times$3 & 197.2$\times$3 \\
J0435+2253 & 2019-12-04 & 267.0$\times$3 & 197.2$\times$3 & 197.2$\times$3 \\
J0508+2422 & 2019-09-08 & 267.0$\times$3 & 197.2$\times$3 & 197.2$\times$3 \\
J064.68+27.97 & 2019-12-15 & 270.0$\times$3 & 197.2$\times$3 & 197.2$\times$3 \\
J065.17+28.17 & 2019-08-01 & 270.0$\times$3 & 197.2$\times$3 & 197.2$\times$3 \\
J0431+2322 & 2019-12-13 & 267.0$\times$3 & 197.2$\times$3 & 197.2$\times$3 \\
J0433+1758 & 2019-08-31 & 267.0$\times$3 & 197.2$\times$3 & 197.2$\times$3 \\
J061.87+23.11 & 2019-08-01 & 267.0$\times$3 & 197.2$\times$3 & 197.2$\times$3 \\
J0409+2000 & 2019-11-25 & 233.0$\times$3 & 197.2$\times$3 & 197.2$\times$3 \\
J075.58+16.94 & 2019-08-29 & 267.0$\times$3 & 197.2$\times$3 & 197.2$\times$3 \\
J059.57+30.63 & 2019-08-23 & 275.0$\times$3 & 197.2$\times$3 & 197.2$\times$3 \\
J0427+2738 & 2019-12-24 & 270.0$\times$3 & 197.2$\times$3 & 197.2$\times$3 \\
J0432+1815 & 2019-09-02 & 267.0$\times$3 & 197.2$\times$3 & 197.2$\times$3 \\
J0433+2251 & 2019-12-14 & 267.0$\times$3 & 197.2$\times$3 & 197.2$\times$3 \\
J0452+3037 & 2019-08-26 & 275.0$\times$3 & 197.2$\times$3 & 197.2$\times$3 \\
J0421+2706 & 2019-12-23 & 270.0$\times$3 & 197.2$\times$3 & 197.2$\times$3 \\
J0421+2754 & 2019-09-02 & 270.0$\times$3 & 197.2$\times$3 & 197.2$\times$3 \\
J0422+1523 & 2019-08-29 & 267.0$\times$3 & 197.2$\times$3 & 197.2$\times$3 \\
J064.68+27.97 & 2019-12-23 & 270.0$\times$3 & 197.2$\times$3 & 197.2$\times$3 \\
J0439+2332 & 2019-08-31 & 267.0$\times$3 & 197.2$\times$3 & 197.2$\times$3 \\
J0452+3037 & 2019-08-26 & 275.0$\times$3 & 197.2$\times$3 & 197.2$\times$3 \\
J0422+1536 & 2019-07-24 & 267.0$\times$3 & 197.2$\times$3 & 197.2$\times$3 \\
J065.88+19.83 & 2019-07-25 & 267.0$\times$3 & 197.2$\times$3 & 197.2$\times$3 \\
J0449+2911 & 2019-08-28 & 270.0$\times$3 & 197.2$\times$3 & 197.2$\times$3 \\
J0421+2706 & 2020-03-01 & 270.0$\times$3 & 197.2$\times$3 & 197.2$\times$3 \\
\hline
\end{tabular}
\label{tab:taurus_obslog}
\end{table*}

The reason why we chose these three filters for observations is that we can build a reddening-insensitive index ($Q_{\mathrm{HST}}$) with the fluxes in the three filters, which is an HST equivalent of Eq.~\ref{eq_Q}.
This can be used to distinguish companion candidates from background stars in just one epoch. 
The technique is detailed in \cite{Dubber2021} and the expression for $Q_{\mathrm{HST}}$ is:
\begin{equation}
Q_{\mathrm{HST}} = -2.5 \log \left( \frac{F_{\mathrm{F850LP}}}{F_{\mathrm{F127M}}} \right) + e \times 2.5 \log \left( \frac{F_{\mathrm{F127M}}}{F_{\mathrm{F139M}}} \right),
\end{equation}
where $e$ is a constant of 5.81 calculated by the extinction ratio in the three bands: $e = (A_{F850LP} - A_{F127M})/(A_{F139M} - A_{F127M})$, adopting R$_V$ = 3.1 from \cite{Fitzpatrick1999}.
The F139M filter can explore the water absorption band at 1.4\,$\mu$m where late-type objects will have strong absorption, while background stars earlier than M6 will not.
Combining results from the three filters, a background star will have $Q_{\mathrm{HST}}$ $\approx$ 0, while an object with a spectral type of late-M, L or T will have $Q_{\mathrm{HST}}$ $>$ 1. 
Candidate companions around the targets should have $Q_{\mathrm{HST}}$ greater than that of the primary sources.
Therefore, this technique can help to select bona fide companions and reject background stellar interlopers.

\section{Photometric analysis}
\label{sec:taurus_phot}
The raw data were calibrated with the automated calibration pipelines provided by the Space Telescope Science Institute (STScI). 
The \textit{\_flt} files are calibrated individual exposures after bias subtraction, dark current subtraction, and flat-field correction.
For the F850LP filter, further 
charge transfer efficiency (CTE) corrections were applied to the \textit{\_flt} files to mitigate the effects of charge transfer inefficiency in CCD detectors, resulting in \textit{\_flc} files.
There are also \textit{\_drz} or \textit{\_drc} (applicable for F850LP) files available that are geometrically corrected and cosmic-ray-cleaned combined images from multiple exposures using the DrizzlePac software \citep{Gonzaga2012}.

\subsection{Aperture photometry}
We used \textsc{photutils} \citep{larry_bradley_2024_13989456} to perform aperture photometry on \textit{\_drz} files for F127M and F139M and \textit{\_drc} files for F850LP. 
We adopted an aperture radius of 0.4" which covers more than 80\% of the energy while limiting the background uncertainty.
The uncertainty is estimated by assuming the Poisson noise for the source flux and also includes the background noise by taking the three-sigma clipped standard deviation of 1000 random apertures on source-free positions, the same as the method used in \cite{Schneider2015,Fontanive2018}.
Then the fluxes of the sources were corrected to an infinite aperture using the encircled energy tables and calibrated to the zero point of the Vega magnitude system using the instructions provided by STScI\footnote{\url{https://www.stsci.edu/hst/instrumentation/wfc3/data-analysis/photometric-calibration}}.
We adopted the typical flux uncertainty of HST photometric calibration: 1.8\% for WFC/UVIS and 2.0\% for WFC/IR.

\subsection{Point Spread Function (PSF) fitting}
\label{sec:taurus_psffit}
We performed \textsc{HST1PASS} on individual \textit{\_flt} or \textit{\_flc} files to search for and detect sources by PSF fitting. \textsc{HST1PASS} is a package written in \textsc{FORTRAN} that uses empirical spatially variable PSFs libraries to run a single pass of source detection in individual exposures \citep{Anderson2022}. 
We adopted a template PSF size of 5\,pixels $\times$ 5\,pixels.
The output products include single-exposure catalogues of positions, fluxes and residuals after the subtraction of detected sources, as well as the derivation of PSF models obtained by perturbing the empirical PSF libraries. 
As we have exposures taken at three dithered positions, for each target, we built six-parameter linear transformations based on well-measured unsaturated sources to establish a reference coordinate system, which maps detected sources in the dithered images to this system.

We further used \textsc{KS2}, a package written in \textsc{FORTRAN} based on \textsc{kitchen-sync} \citep{Anderson2008, Bellini2017, Nardiello2018}, to perform source finding and PSF-fitting simultaneously in multiple exposures using the transformations and perturbed PSFs. 
Since it combines multiple exposures, \textsc{KS2} can detect faint sources that may not be detectable in single exposures.
\rev{There are three important PSF fitting parameters of \textsc{KS2} \cite{Bedin2008, Scalco2021}.
The parameter `o' represents the initial ratio (before subtracting neighbouring sources) of the light from nearby neighbours within the fitting radius to the light from the target star.
The parameter quality-of-fit (QFIT) indicates how well the PSF-fitting process performed when measuring a star's position and brightness with values closer to one suggesting that the source closely matches the assumed PSF model.
The parameter RADXS quantifies deviations from the expected PSF shape by comparing the observed flux just outside the PSF core to that predicted by the PSF model. Extended sources like galaxies typically show large positive RADXS values, whereas unusually sharp sources, such as cosmic rays or hot pixels, yield large negative values.}

The data reduction process follows the procedures described in \cite{Scalco2021, Fontanive2023}.
We supersampled the stacked images by a factor of 2. 
The output products also contain the stacked source-subtracted residuals.
The current version of \textsc{KS2} is only configured to fully reduce data in WFC3/IR filters.
Therefore, we had second-pass photometry for sources in the two IR filters but only stacked images and source-subtracted residuals for F850LP. 
This is enough for our companion search because the candidates would be much brighter in F127M than F850LP as they are planetary-mass objects and emit most of their energy in the infrared, though angular resolution is worse than in F850LP.
As a result, we have photometric measurements from \textsc{KS2} only for the two IR filters.
The instrumental magnitudes of sources were measured by PSF fitting in \textsc{KS2}.
We followed the same method demonstrated in \cite{Scalco2021} to perform the photometric calibration, which includes corrections for the difference between PSF fitting and aperture photometry as well as the zero point (ZP) of the Vega magnitude system.
The adopted ZP is 23.649\,mag in F127M and 23.367\,mag in F139M.

\begin{figure}
    \centering
    \includegraphics[width=1\columnwidth]{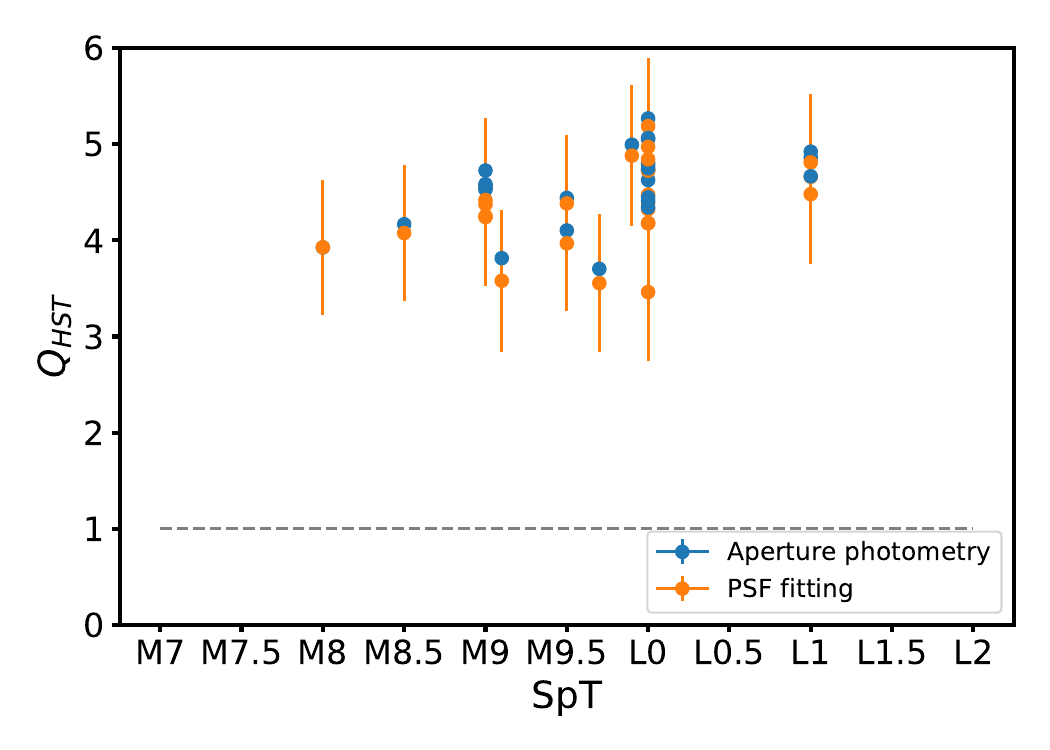}
    \caption[$Q_\mathrm{HST}$ of primary sources]{$Q_\mathrm{HST}$ of primary sources calculated by aperture photometry and PSF fitting. The uncertainty of SpT is on the order of 1.5. All the targets have $Q_\mathrm{HST}$ $>$ 1, verifying that they have SpT later than mid-M. Stars with earlier SpT should have $Q_\mathrm{HST}$ $<$ 1.}
    \label{fig:taurus_Q}
\end{figure}

Table~\ref{tab:taurus_hstphotometry} lists the photometric measurements of aperture photometry and PSF fitting of the targets. The photometric measurements of the two methods agree with each other, although the uncertainty of aperture photometry is likely underestimated.
We also computed the $Q_\mathrm{HST}$ using the photometric measurements of the two methods. We used the photometric flux measured by aperture photometry in F850LP for the $Q_\mathrm{HST}$ calculation of \textsc{KS2} as it only has measurements in the two IR filters.
The resulting $Q_\mathrm{HST}$ by aperture photometry has a mean offset of 0.2 higher than the one by PSF fitting.
All the targets have $Q_\mathrm{HST}$ $>$ 1 as expected, as shown in Fig.~\ref{fig:taurus_Q}.

\begin{table*}
\centering
\caption[HST photometry of the targets]{Photometric results of the targets.}
\label{tab:taurus_hstphotometry}
\begin{tabular}{cccccccc}
\hline\hline
Name & \multicolumn{4}{c}{Aperture photometry} & \multicolumn{3}{c}{PSF fitting} \\
\cmidrule(lr){2-5} \cmidrule(lr){6-8}
 & F850LP (mag) & F127M (mag) & F139M (mag) & $Q_\mathrm{HST}$ & F127M (mag) & F139M (mag) & $Q_\mathrm{HST}$\\
\hline
J0435+2307 & 19.13$\pm$0.02 & 17.14$\pm$0.02 & 17.45$\pm$0.02 & 4.86$\pm$0.20 & 17.13$\pm$0.09 & 17.40$\pm$0.06 & 4.66$\pm$0.71 \\
J0401+2557 & 18.46$\pm$0.02 & 16.68$\pm$0.02 & 17.03$\pm$0.02 & 4.92$\pm$0.20 & 16.68$\pm$0.09 & 17.01$\pm$0.06 & 4.81$\pm$0.71 \\
J0424+2641 & 19.49$\pm$0.03 & 17.36$\pm$0.02 & 17.58$\pm$0.02 & 4.53$\pm$0.20 & 17.36$\pm$0.09 & 17.56$\pm$0.06 & 4.37$\pm$0.71 \\
J0422+2655 & 20.30$\pm$0.04 & 17.82$\pm$0.02 & 18.00$\pm$0.02 & 4.67$\pm$0.21 & 17.83$\pm$0.09 & 17.98$\pm$0.06 & 4.48$\pm$0.72 \\
J0422+1947 & 18.02$\pm$0.02 & 16.34$\pm$0.02 & 16.65$\pm$0.02 & 4.58$\pm$0.20 & 16.37$\pm$0.09 & 16.63$\pm$0.06 & 4.25$\pm$0.70 \\
J0435+2253 & 19.38$\pm$0.03 & 17.20$\pm$0.02 & 17.45$\pm$0.02 & 4.72$\pm$0.20 & 17.20$\pm$0.09 & 17.40$\pm$0.06 & 4.47$\pm$0.72 \\
J0508+2422 & 18.57$\pm$0.02 & 16.86$\pm$0.02 & 17.18$\pm$0.02 & 4.73$\pm$0.20 & 16.86$\pm$0.09 & 17.16$\pm$0.07 & 4.56$\pm$0.71 \\
J064.68+27.97 & 18.21$\pm$0.02 & 16.42$\pm$0.02 & 16.75$\pm$0.02 & 4.81$\pm$0.20 & 16.42$\pm$0.09 & 16.74$\pm$0.06 & 4.74$\pm$0.71 \\
J065.17+28.17 & 18.33$\pm$0.02 & 16.52$\pm$0.02 & 16.89$\pm$0.02 & 5.05$\pm$0.20 & 16.53$\pm$0.09 & 16.78$\pm$0.06 & 4.38$\pm$0.71 \\
J0431+2322 & 18.97$\pm$0.02 & 16.69$\pm$0.02 & 16.87$\pm$0.02 & 4.41$\pm$0.20 & 16.70$\pm$0.09 & 16.84$\pm$0.06 & 4.18$\pm$0.72 \\
J0433+1758 & 19.59$\pm$0.03 & 17.53$\pm$0.02 & 17.81$\pm$0.02 & 4.78$\pm$0.20 & 17.53$\pm$0.09 & 17.73$\pm$0.06 & 4.32$\pm$0.72 \\
J061.87+23.11 & 18.17$\pm$0.02 & 16.37$\pm$0.02 & 16.73$\pm$0.02 & 4.99$\pm$0.20 & 16.37$\pm$0.09 & 16.71$\pm$0.07 & 4.88$\pm$0.73 \\
J0409+2000 & 16.78$\pm$0.02 & 15.43$\pm$0.02 & 15.66$\pm$0.02 & 3.81$\pm$0.20 & 15.39$\pm$0.09 & 15.58$\pm$0.06 & 3.58$\pm$0.74 \\
J075.58+16.94 & 17.32$\pm$0.02 & 15.95$\pm$0.02 & 16.23$\pm$0.02 & 4.10$\pm$0.20 & 15.95$\pm$0.09 & 16.20$\pm$0.06 & 3.97$\pm$0.71 \\
J059.57+30.63 & 17.32$\pm$0.02 & 15.87$\pm$0.02 & 16.07$\pm$0.02 & 3.70$\pm$0.20 & 15.86$\pm$0.09 & 16.03$\pm$0.07 & 3.56$\pm$0.72 \\
J0427+2738 & 17.91$\pm$0.02 & 16.01$\pm$0.02 & 16.26$\pm$0.02 & 4.44$\pm$0.20 & 16.01$\pm$0.09 & 16.25$\pm$0.06 & 4.38$\pm$0.71 \\
J0432+1815 & 18.95$\pm$0.02 & 16.82$\pm$0.02 & 17.06$\pm$0.02 & 4.63$\pm$0.20 & 16.82$\pm$0.09 & 17.07$\pm$0.06 & 4.73$\pm$0.72 \\
J0433+2251 & 19.99$\pm$0.03 & 17.75$\pm$0.02 & 17.95$\pm$0.02 & 4.45$\pm$0.21 & 17.77$\pm$0.09 & 17.92$\pm$0.07 & 4.18$\pm$0.74 \\
J0452+3037 & 18.83$\pm$0.02 & 16.89$\pm$0.02 & 17.27$\pm$0.02 & 5.27$\pm$0.20 & 16.88$\pm$0.09 & 17.25$\pm$0.06 & 5.19$\pm$0.71 \\
J0421+2706 & 18.92$\pm$0.02 & 16.94$\pm$0.02 & 17.19$\pm$0.02 & 4.54$\pm$0.20 & 16.96$\pm$0.09 & 17.17$\pm$0.06 & 4.24$\pm$0.72 \\
J0421+2754 & 18.54$\pm$0.02 & 16.47$\pm$0.02 & 16.80$\pm$0.02 & 5.06$\pm$0.20 & 16.48$\pm$0.09 & 16.79$\pm$0.06 & 4.97$\pm$0.71 \\
J0422+1523 & 19.95$\pm$0.03 & 17.88$\pm$0.02 & 18.15$\pm$0.02 & 4.75$\pm$0.21 & 17.87$\pm$0.09 & 18.15$\pm$0.06 & 4.84$\pm$0.70 \\
J0439+2332 & 18.98$\pm$0.02 & 17.03$\pm$0.02 & 17.29$\pm$0.02 & 4.57$\pm$0.20 & 17.03$\pm$0.09 & 17.27$\pm$0.06 & 4.42$\pm$0.70 \\
J0422+1536 & 17.14$\pm$0.02 & 15.71$\pm$0.02 & 15.95$\pm$0.02 & 3.93$\pm$0.20 & 15.70$\pm$0.09 & 15.93$\pm$0.06 & 3.93$\pm$0.70 \\
J065.88+19.83 & 14.84$\pm$0.02 & 13.45$\pm$0.02 & 13.77$\pm$0.02 & 4.34$\pm$0.20 & 13.44$\pm$0.09 & 13.60$\pm$0.06 & 3.46$\pm$0.72 \\
J0449+2911 & 16.07$\pm$0.02 & 14.72$\pm$0.02 & 15.02$\pm$0.02 & 4.17$\pm$0.20 & 14.73$\pm$0.09 & 15.01$\pm$0.06 & 4.08$\pm$0.70 \\
\hline
\end{tabular}
\end{table*}

\subsection{Primary masses}
For targets without Two Micron All Sky Survey (2MASS) \citep{Skrutskie2006} $J$ band measurements, we converted from the \rev{Mauna Kea Observatories} (MKO) photometric system \citep{Simons2002,Tokunaga2002} to the 2MASS photometric system using the relation $J_\mathrm{2MASS}$ = $J_\mathrm{MKO}$ + 0.05 which is for dwarfs with spectral types later than M6 \citep{ZhangZ2018}.
Then we converted apparent $J_\mathrm{2MASS}$ magnitudes to absolute magnitudes $M_{J\mathrm{2MASS}}$ using the distance listed in Table~\ref{tab:taurus_target_info} and extinction listed in Table~\ref{tab:taurus_mass}.
The $A_V$ was converted to $A_J$ by $A_J$ = 0.267 $\times$ $A_V$ \citep{Fitzpatrick1999}.
We estimated the bolometric correction $BC_J$ by SpT (with uncertainty of 1.5 subtype) using the polynomial relation provided in \cite{Filippazzo2015}.
The bolometric luminosity $Lbol$ is calculated by 
\begin{equation}
  Lbol = \log_{10}\left(\frac{L_{\mathrm{bol}}}{L_{\odot}}\right) = -\frac{M_{J\mathrm{2MASS}} + BC_J - M_{\mathrm{bol},\odot}}{2.5},
\end{equation}
where $M_{\mathrm{bol},\odot}$ is taken as 4.73 mag \citep{Torres2010}.

Then we chose the AMES-Dusty2000 evolutionary models \citep{Chabrier2000} to derive the mass. We used the rejection sampling method \citep[e.g.,][]{Dupuy2017} to evaluate the mass and $T_\mathrm{eff}$ by the derived bolometric luminosity and age.
The results are listed in Table~\ref{tab:taurus_mass} together with the 2MASS and MKO magnitudes of the targets.
Most of the targets have masses $<$ 20\,\mj{} with three targets lower than 10\,\mj{}. J0508+2422 is the object of the lowest mass in this sample, which is 6.29$_{-1.38}^{+1.49}$\,\mj{}.
There is one problematic target J065.88+19.83: it has an estimated mass of $\sim$34\,\mj{}. This mass is overestimated because its actual age should be younger than 1\,Myr \citep{ZhangZ2018}.
Because AMES-Dusty2000 evolutionary models only provide values for objects $\geq$ 1\,Myr, it estimates a heavier mass when the age is older than the actual age given the same luminosity.
Constraining the parameters of objects younger than 1\,Myr is challenging as all the current evolutionary models have large uncertainties at this range.
Hence, we assumed an age of 1\,Myr to estimate its mass because this is the youngest age the AMES-Dusty2000 models provided. 
All the targets have $T_\mathrm{eff}$ over 1800\,K, typical $T_\mathrm{eff}$ for warm and dusty brown dwarfs.

\begin{landscape}
\begin{table}
\centering
\caption[Photometry and mass of the targets]{Photometry and mass of the targets.}
\label{tab:taurus_mass}
\begin{tabular}{cccccccccc}
\hline\hline
Name & $J_\mathrm{2MASS}$ (mag) & $H_\mathrm{2MASS}$ (mag) & $K_{2MASS}$ (mag) & $J_\mathrm{MKO}$ (mag) & $A_V$ (mag) & Lbol (dex) & $T_\mathrm{eff}$ (K) & Mass$^{a}$ (\mj{}) & Mass$^{b}$ (\mj{})\\
& & & & & & & & (Dusty) & (ATMO)\\
\hline
J0435+2307 & --- & --- & --- & 17.27$\pm$0.04 & 2.33 & -3.13$\pm$0.17 & 2096$_{-241}^{+133}$ & 12.43$_{-4.39}^{+3.20}$ & 11.18 \\
J0401+2557 & 16.81$\pm$0.17 & 15.73$\pm$0.14 & 15.36$\pm$0.17 & --- & 0.00 & -3.17$\pm$0.18 & 1978$_{-236}^{+122}$ & 6.80$_{-2.04}^{+2.04}$ & 5.41 \\
J0424+2641 & --- & --- & --- & 17.34$\pm$0.09 & 3.04 & -3.20$\pm$0.13 & 2061$_{-147}^{+105}$ & 12.21$_{-3.73}^{+2.54}$ & 11.03 \\
J0422+2655 & --- & --- & --- & 17.94$\pm$0.11 & 5.21 & -3.09$\pm$0.18 & 2121$_{-252}^{+135}$ & 12.81$_{-4.57}^{+3.31}$ & 11.73 \\
J0422+1947 & 16.63$\pm$0.13 & 15.88$\pm$0.17 & 15.12$\pm$0.16 & 16.44$\pm$0.03 & 0.95 & -3.12$\pm$0.14 & 2264$_{-167}^{+157}$ & 24.25$_{-8.46}^{+9.88}$ & 15.28 \\
J0435+2253 & --- & --- & --- & 17.45$\pm$0.07 & 4.25 & -3.07$\pm$0.15 & 2149$_{-190}^{+119}$ & 13.40$_{-4.35}^{+2.99}$ & 13.56 \\
J0508+2422 & 16.93$\pm$0.14 & 16.47$\pm$0.24 & 15.82$\pm$0.22 & 16.99$\pm$0.03 & 0.75 & -3.26$\pm$0.14 & 1922$_{-135}^{+103}$ & 6.29$_{-1.39}^{+1.49}$ & 5.48 \\
J064.68+27.97 & 16.69$\pm$0.12 & 15.85$\pm$0.12 & 15.25$\pm$0.12 & 16.64$\pm$0.03 & 1.08 & -3.08$\pm$0.16 & 2137$_{-204}^{+122}$ & 13.18$_{-4.38}^{+3.03}$ & 12.98 \\
J065.17+28.17 & 16.77$\pm$0.12 & 15.98$\pm$0.15 & 15.23$\pm$0.11 & 16.67$\pm$0.02 & 0.79 & -3.15$\pm$0.16 & 2090$_{-206}^{+123}$ & 12.43$_{-4.16}^{+2.97}$ & 12.06 \\
J0431+2322 & 16.62$\pm$0.16 & 15.78$\pm$0.18 & 14.71$\pm$0.11 & 16.97$\pm$0.06 & 0.13 & -3.16$\pm$0.16 & 2079$_{-223}^{+129}$ & 12.22$_{-4.21}^{+3.07}$ & 10.31 \\
J0433+1758 & --- & --- & --- & 17.59$\pm$0.07 & 0.84 & -3.49$\pm$0.13 & 1861$_{-143}^{+102}$ & 10.44$_{-2.11}^{+1.77}$ & 9.03 \\
J061.87+23.11 & 16.41$\pm$0.12 & 15.70$\pm$0.15 & 15.19$\pm$0.13 & --- & 1.68 & -2.91$\pm$0.18 & 2240$_{-258}^{+143}$ & 14.95$_{-5.33}^{+3.52}$ & 14.17 \\
J0409+2000 & 15.57$\pm$0.06 & 14.93$\pm$0.07 & 14.38$\pm$0.05 & --- & 0.21 & -2.89$\pm$0.16 & 2269$_{-195}^{+127}$ & 15.60$_{-5.03}^{+3.16}$ & 15.57 \\
J075.58+16.94 & 16.14$\pm$0.08 & 15.37$\pm$0.08 & 14.74$\pm$0.09 & --- & 0.00 & -3.01$\pm$0.17 & 2354$_{-185}^{+169}$ & 25.01$_{-7.98}^{+10.06}$ & 16.38 \\
J059.57+30.63 & 16.02$\pm$0.08 & 15.37$\pm$0.09 & 14.72$\pm$0.09 & --- & 0.00 & -2.95$\pm$0.17 & 2411$_{-218}^{+167}$ & 27.35$_{-9.48}^{+10.86}$ & 19.54 \\
J0427+2738 & 16.10$\pm$0.09 & 15.44$\pm$0.11 & 14.64$\pm$0.09 & --- & 1.03 & -2.88$\pm$0.14 & 2285$_{-155}^{+114}$ & 16.05$_{-4.65}^{+2.80}$ & 15.11 \\
J0432+1815 & --- & 16.02$\pm$0.24 & 15.02$\pm$0.12 & 17.02$\pm$0.06 & 3.91 & -2.93$\pm$0.15 & 2436$_{-227}^{+167}$ & 31.18$_{-13.29}^{+11.76}$ & 19.86 \\
J0433+2251 & --- & --- & --- & 17.71$\pm$0.11 & 2.08 & -3.41$\pm$0.16 & 1892$_{-207}^{+128}$ & 9.94$_{-3.44}^{+2.71}$ & 8.29 \\
J0452+3037 & --- & --- & 13.84$\pm$0.04 & 17.08$\pm$0.05 & 0.88 & -3.20$\pm$0.12 & 2049$_{-131}^{+86}$ & 10.04$_{-2.10}^{+2.02}$ & 8.94 \\
J0421+2706 & --- & 16.12$\pm$0.15 & 15.32$\pm$0.14 & 17.15$\pm$0.06 & 2.67 & -3.16$\pm$0.13 & 2095$_{-142}^{+101}$ & 12.67$_{-3.79}^{+2.51}$ & 12.49 \\
J0421+2754 & 16.73$\pm$0.12 & 15.40$\pm$0.08 & 14.84$\pm$0.09 & 16.57$\pm$0.03 & 2.16 & -2.98$\pm$0.16 & 2207$_{-199}^{+121}$ & 14.37$_{-4.66}^{+3.08}$ & 14.30 \\
J0422+1523 & --- & --- & --- & 18.05$\pm$0.08 & 3.26 & -3.41$\pm$0.15 & 2032$_{-209}^{+116}$ & 16.98$_{-5.32}^{+7.60}$ & 13.46 \\
J0439+2332 & 16.94$\pm$0.15 & 16.30$\pm$0.20 & 15.62$\pm$0.19 & 17.15$\pm$0.05 & 0.00 & -3.44$\pm$0.11 & 1855$_{-101}^{+83}$ & 7.82$_{-1.13}^{+1.09}$ & 6.31 \\
J0422+1536 & --- & --- & --- & 15.69$\pm$0.01 & 0.71 & -2.79$\pm$0.12 & 2562$_{-182}^{+107}$ & 37.16$_{-15.87}^{+10.14}$ & 30.61 \\
J065.88+19.83 & 13.78$\pm$0.02 & 13.03$\pm$0.03 & 12.59$\pm$0.03 & --- & 0.00 & -1.99$\pm$0.16 & 2746$_{-101}^{+55}$ & 36.39$_{-9.59}^{+7.31}$ & 31.01 $^c$ \\
J0449+2911 & --- & --- & --- & 14.87$\pm$0.01 & 0.00 & -2.54$\pm$0.12 & 2555$_{-123}^{+95}$ & 23.08$_{-4.31}^{+5.61}$ & 20.82 \\
\hline
\end{tabular}
\begin{tablenotes}
\item Notes: 2MASS photometric magnitudes are from Simbad. $J_\mathrm{MKO}$ were measured in the WBIS and queries from UKIDSS.
(a).Mass estimated with bolometric luminosity and dusty models.
(b).Mass estimated with magnitude in F127M and ATMO models.
(c).The mass is overestimated because there are no reliable models for objects younger than 1\,Myr.
\end{tablenotes}
\end{table}
\end{landscape}

\section{Multiplicity analysis}
\label{sec:taurus_mul}
\subsection{Companion search}
\begin{figure}
    \centering
    \includegraphics[width=1\columnwidth]{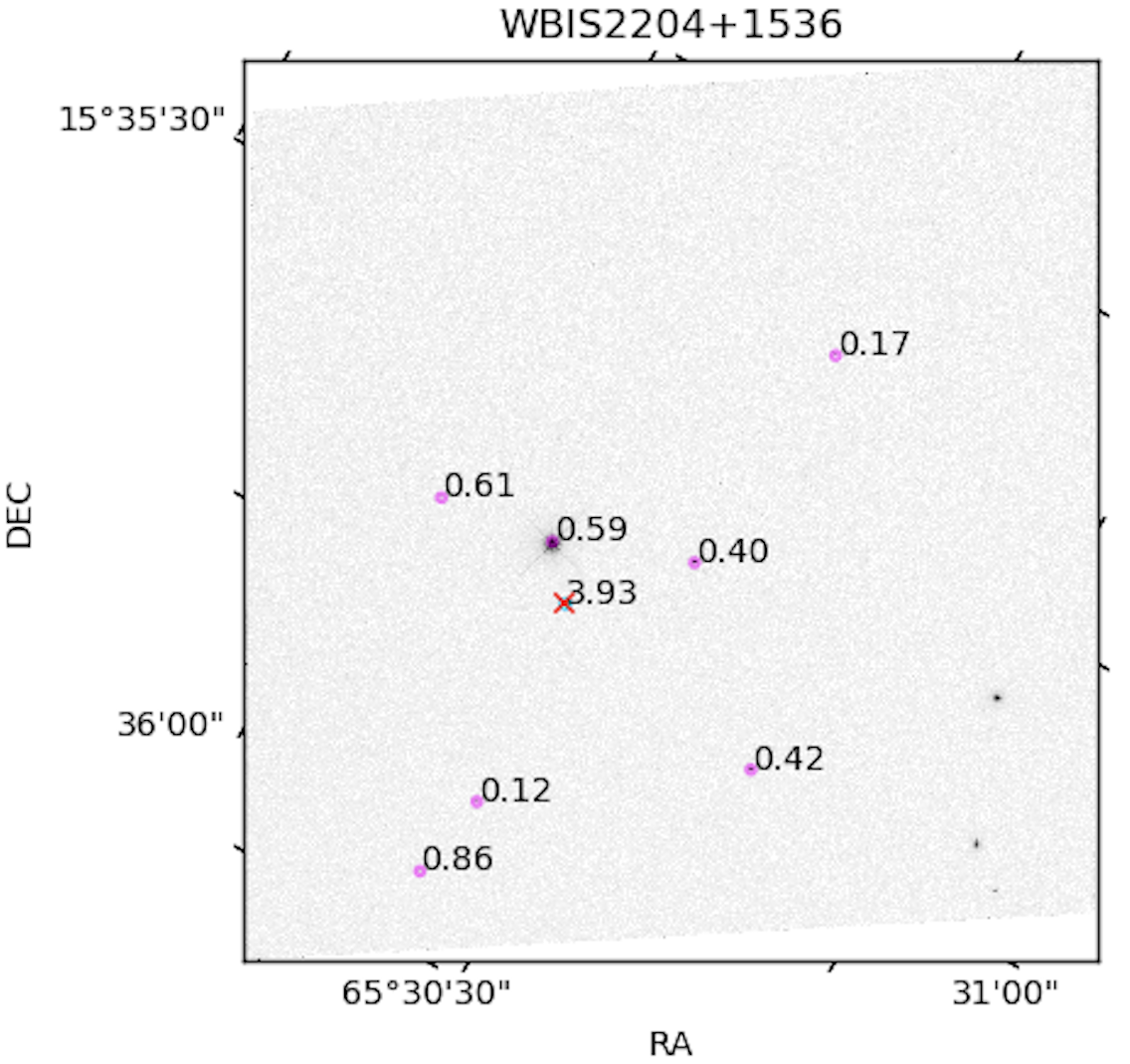}
    \caption[$Q_\mathrm{HST}$ of detected sources around WBIS2204+1536]{$Q_\mathrm{HST}$ of detected sources in the FoV of WBIS2204+1536. The primary target is marked by the red cross and other detected sources are marked by pink circles. Their $Q_\mathrm{HST}$ are labelled adjacent to them. Only the primary target has $Q_\mathrm{HST}$ $>$ 1.}
    \label{fig:taurus_Q_fov}
\end{figure}

We used our $Q_\mathrm{HST}$ technique to search for widely-separated companion candidates with aperture photometry. The source detection was performed with DAOStarFinder \citep{Stetson1987} embedded in \textsc{photutils}. We calculated the $Q_\mathrm{HST}$ of all the detected sources in the FoV of our targets. No candidates have $Q_\mathrm{HST}$ $>$ 1.
Figure~\ref{fig:taurus_Q_fov} shows the FoV of an example target (WBIS2204+1536) in F850LP with $Q_\mathrm{HST}$ of all detected sources.
Only the primary target has $Q_\mathrm{HST}$ over 1, while the other sources have $Q_\mathrm{HST}$ $<$ 1, suggesting that they are background stars.
We also performed a visual inspection of the residual images after subtraction of the primary target and two waves of source finding with \textsc{KS2}. No close-in companion candidates were found.

\subsection{Artificial star tests}
To evaluate the sensitivity and reliability of our source detection method, we conducted artificial star injection and retrieval tests (ASTs) using the same method detailed in \cite{Fontanive2023}. 
For each target, we injected 4 $\times$ 10$^{5}$ artificial sources using the PSFs modelled in Sect.~\ref{sec:taurus_psffit} to all individual frames in F127M and F139M with one source injected and measured at a time.
The artificial sources were uniformly injected across the FoV with a uniform distribution for instrumental magnitude between -7 and 4 mag in F127M and F139M filters, separately, which corresponds to an apparent magnitude of 16.5--27.5 mag in F127M and 16.4--27.4 mag in F139M.

\textsc{KS2} outputs various parameters summarising the PSF fitting results. 
We explored these parameters and determined the criteria to identify if an artificial source is well recovered by comparing its injected and fitted values, following the procedures in \cite{Fontanive2023}. The criteria can then be used to determine if a real source is well detected in a given filter. 
To start with, we applied an initial selection to filter out obvious dubious sources with sources that were retained following the criteria:
\begin{itemize}
    \item sources are detected in all three dithered images;
    \item the measured flux, photometric root mean square (rms), and sky rms are valid numbers ($>$0 and not flagged values (e.g. 9999));
    \item the measured flux is higher than local sky rms;
    \item the absolute difference between the injected and retrieved positions is within 1 pixel in X and Y, respectively;
    \item the absolute difference between the injected and retrieved magnitudes is within $-$2.5 $\times$ log$_{10}$(2).
\end{itemize}

\begin{figure*}
	\includegraphics[width=2\columnwidth]{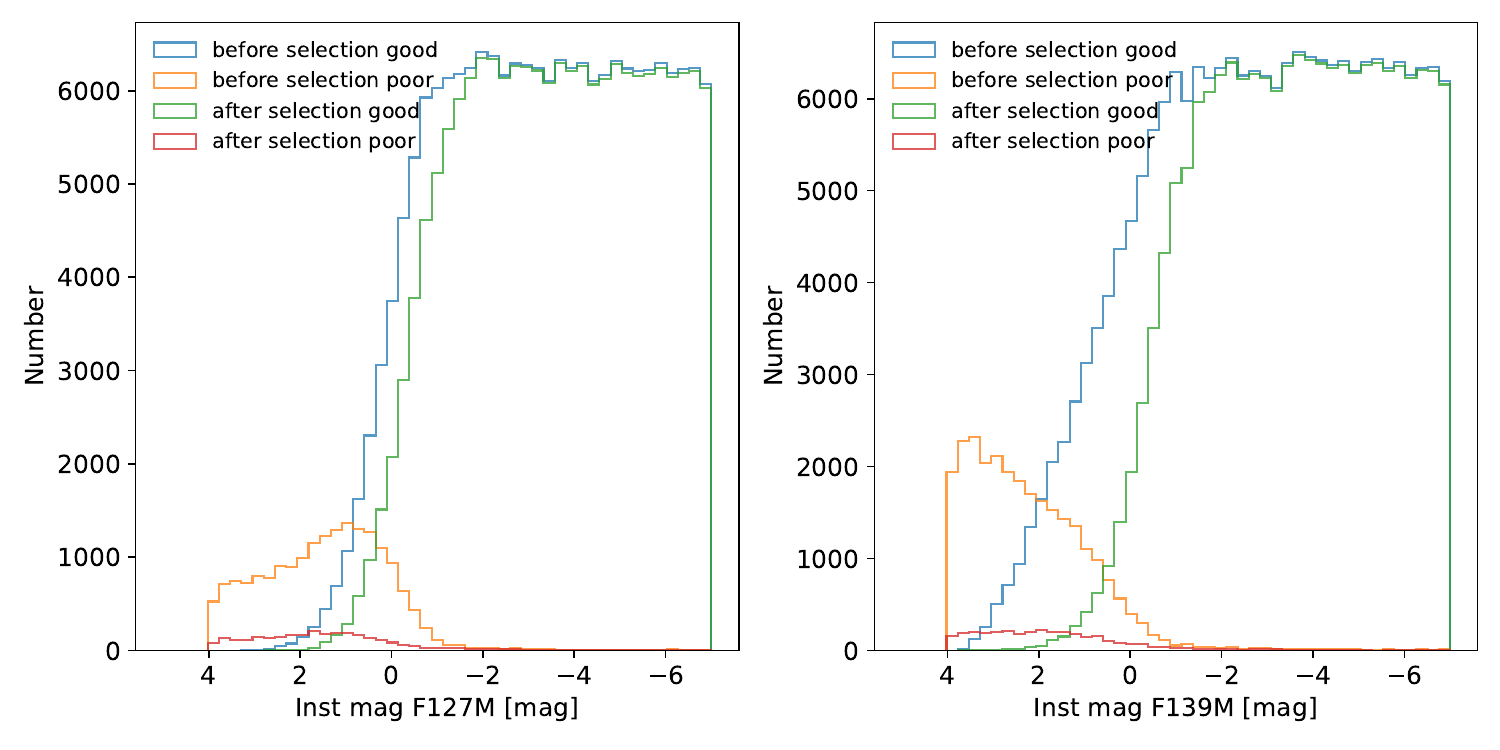}
    \caption[An example of well-recovered and poorly-recovered artificial stars]{An example of well-recovered and poorly-recovered artificial stars before and after applying the selection threshold of `o', QFIT, and RADXS. `Good' and `poor' refer to stars that pass the initial selection and not, separately.}
    \label{fig:taurus_selection_example}
\end{figure*}

The remaining sources after the initial selection are well-detected sources. From here, we made a test of setting the threshold of three PSF fitting parameters \rev{(`o', QFIT, and RADEX)} to retain as many well-recovered sources as possible while excluding as many poorly-recovered sources as possible at the same time. 
We explored the values of `o' in the range of [2, 6], QFIT in the range of [0.7, 0.95], and the absolute value of RADXS in the range of [0.2, 0.5] by grid search to count the number of well-recovered sources and the number of poorly-recovered sources.
The optimised combination of them is: `o' $<$ 4, QFIT $>$ 0.9, and $-$0.35 $<$ RADXS $<$ 0.35. We applied the selection threshold independently in F127M and F139M.
Figure~\ref{fig:taurus_selection_example} shows an example of the number of artificial sources before and after our selection based on the three parameters. 
Poorly-measured sources are mainly sources with instrumental magnitudes fainter than $-$1\,mag.
Less than 15\% of the poorly-measured sources passed the selection threshold.
Over 80\% of the sources that passed the initial criteria are retained after applying the selection threshold, and over 99\% for sources brighter than $-$1\,mag are retained. 

\subsection{Survey completeness}
In order to reach a robust statistical result for the binary fraction for any surveys, it is vital to estimate the completeness of the survey, which evaluates how likely we can detect a companion at a given separation and contrast. 
A robust way to estimate completeness is to perform artificial star injections, as we did in the previous section. 
As we search for companions of the primary brown dwarfs, we injected 2 $\times$ 10$^6$ artificial sources as a function of the instrumental magnitude and separation to the primary targets with uniform distributions in instrumental magnitude, radial separation, and phase angle. 
As in \cite{Fontanive2023}, we split the injection into four sets: 
\begin{enumerate}
    \item 4 $\times$ 10$^5$ sources of instrumental magnitude in the range of [-7, 4] mag injected between 1 and 20 pixels from the primary target;
    \item 6 $\times$ 10$^5$ sources of instrumental magnitude in the range of [-3, 2] mag injected between 1 and 20 pixels from the primary target;
    \item 4 $\times$ 10$^5$ sources of instrumental magnitude in the range of [-7, 4] mag injected beyond 20 pixels from the primary target to the edge of the image;
    \item 6 $\times$ 10$^5$ sources of instrumental magnitude in the range of [-3, 2] mag injected beyond 20 pixels from the primary target to the edge of the image.
\end{enumerate}

This division enabled us to fully explore the entire image while focusing on the parameter space where potential companions are likely located: slightly fainter than and at a small separation to the primary target.
The injections were performed in both F127M and F139M filters.
As the F127M filter has a smaller PSF size which results in better sensitivity at smaller separations, we took the ASTs in F127M to represent the completeness of our survey.
We conducted the ASTs of the two million sources for each target and applied the initial selection and selection threshold to them.
Then we determined the grid size of the completeness maps: an interval of 1 pixel for separation smaller than 20 pixels, an interval of 20 pixels for separation larger than 20 pixels, an interval of 0.5\,mag for magnitude in the range of [-7, -3) and [2, 4], and an interval of 0.1\,mag for magnitude in the range of [-3, 2).
The detection probability of each grid is then calculated by the fraction of artificial stars that passed the initial selection and selection threshold among all the artificial stars injected in that grid.
For separations where only a fraction of the area of the annuli centred on the primary target is located inside the image, we need to multiply the detection probability of these separations by that fraction. Otherwise, the detection probability at these wide separations would be overestimated because we can only detect companions that are located inside the image.

\begin{figure}
    \includegraphics[width=1\columnwidth]{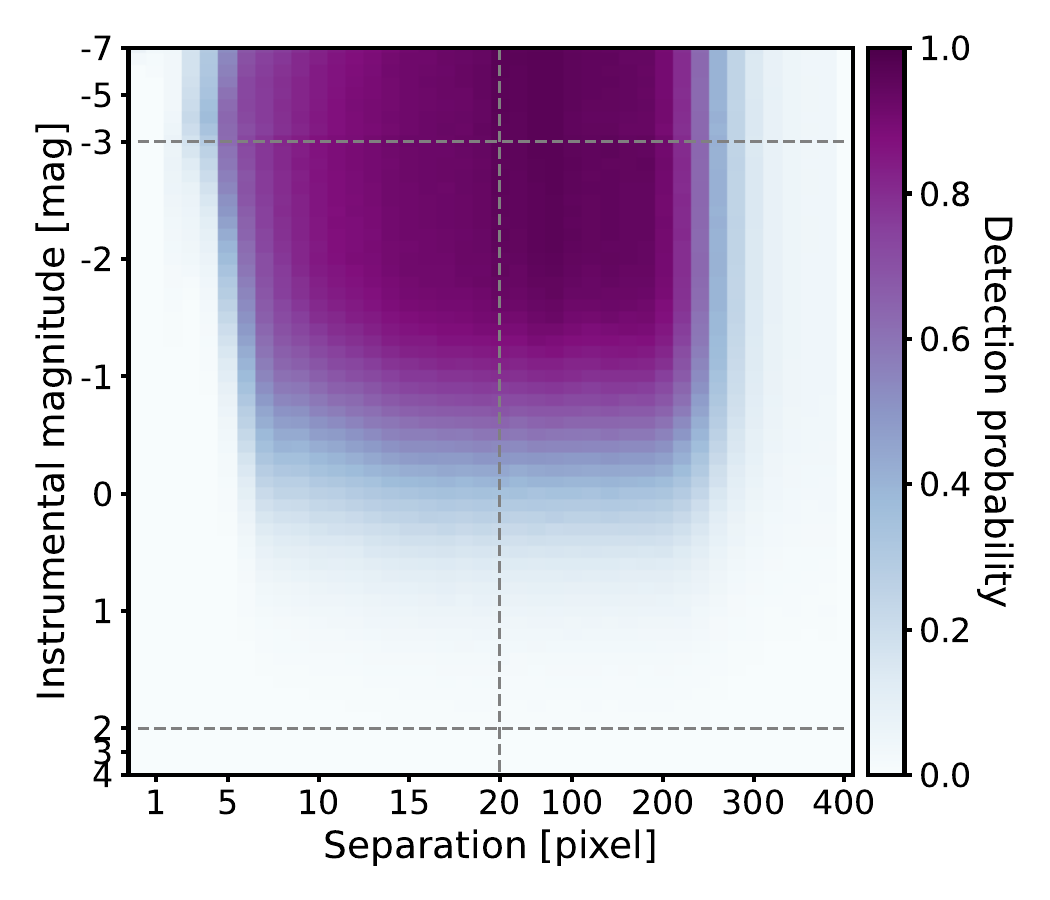}
    \caption[Average completeness map of this survey]{Average completeness map of this survey in F127M.}
    \label{fig:taurus_average_dectmap}
\end{figure}

The observation setup and magnitude of most targets are the same, and therefore the completeness maps of most targets are similar, though some bright targets have shallower detection at smaller separations.
For targets that have multiple observations, we chose the observation that has the best completeness map for each target.
The average completeness map of the 26 targets in this survey is presented in Fig.~\ref{fig:taurus_average_dectmap}. We can detect companions with instrumental magnitudes brighter than $\sim$-1\,mag at separations beyond $\sim$5 pixels with a probability higher than 50\%. 
At separations within three pixels, the detection probability is lower than 20\%. The decrease in the detection probability at wide separations is due to the fact that only parts of the annuli are covered by our observations there and that the PSF fitting becomes less accurate when only a fraction of the source's PSF lies inside the image.

We used the pixel scale and photometric zero point to convert the completeness map as a function of pixel position and instrumental magnitude to a function of apparent magnitude and separation in au. 
While dusty models such as AMES-Dusty2000 are suitable for warm brown dwarfs with effective temperature \teff{} over $\sim$2000\,K, the potential companions probed by this survey are expected to have masses smaller than the masses of the primaries, e.g. $<$ 20\,\mj{} and even $<$ 1\,\mj{}. For those low-mass objects, ATMO models are more suitable, and they are also one of the few models that provide photometry and mass in the F127M filter.
We therefore used the ATMO evolutionary models of chemical equilibrium \citep{Chabrier2023} to estimate mass by magnitude in the F127M filter.
ATMO models are less accurate for warm brown dwarfs because they miss some opacity in the atmosphere when \teff{} is higher than 2000\,K \citep{Phillips2020}.
We compared the mass of the primary target estimated by F127M with ATMO to the mass estimated by bolometric luminosity with AMES-Dusty2000. 
While ATMO produces smaller mass estimates than AMES-Dusty2000, the differences are within 1$\sigma$ for most targets except for three targets: J0422+1947 (1.1$\sigma$), J075.58+16.94 (1.1$\sigma$), and J0439+2332 (1.3$\sigma$).
Given that the differences are small, we adopted the primary mass estimated by F127M magnitude with ATMO models in the conversion of the completeness map for consistency.
These masses are also presented in Table~\ref{tab:taurus_mass}.

\begin{figure*}
    \includegraphics[width=1\columnwidth]{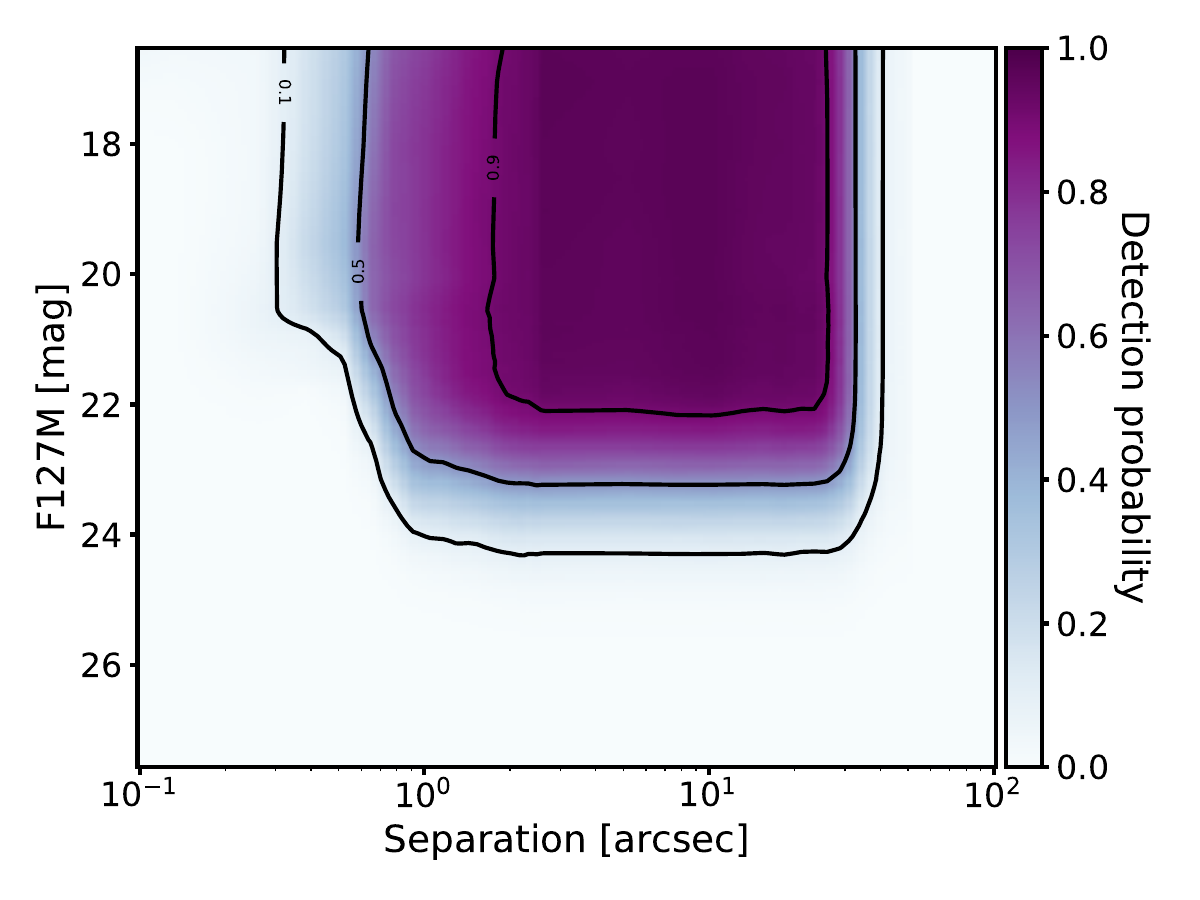}
    \includegraphics[width=1\columnwidth]{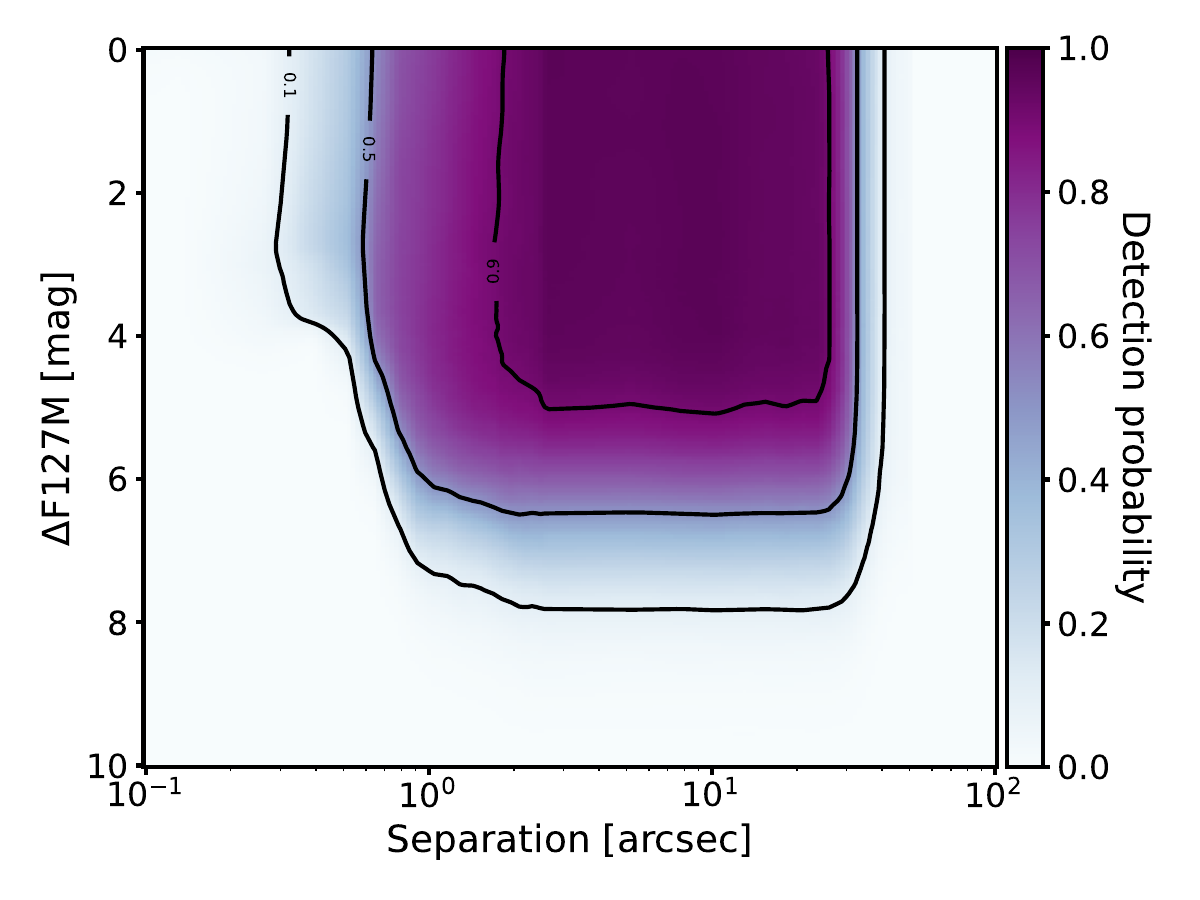}
    \includegraphics[width=1\columnwidth]{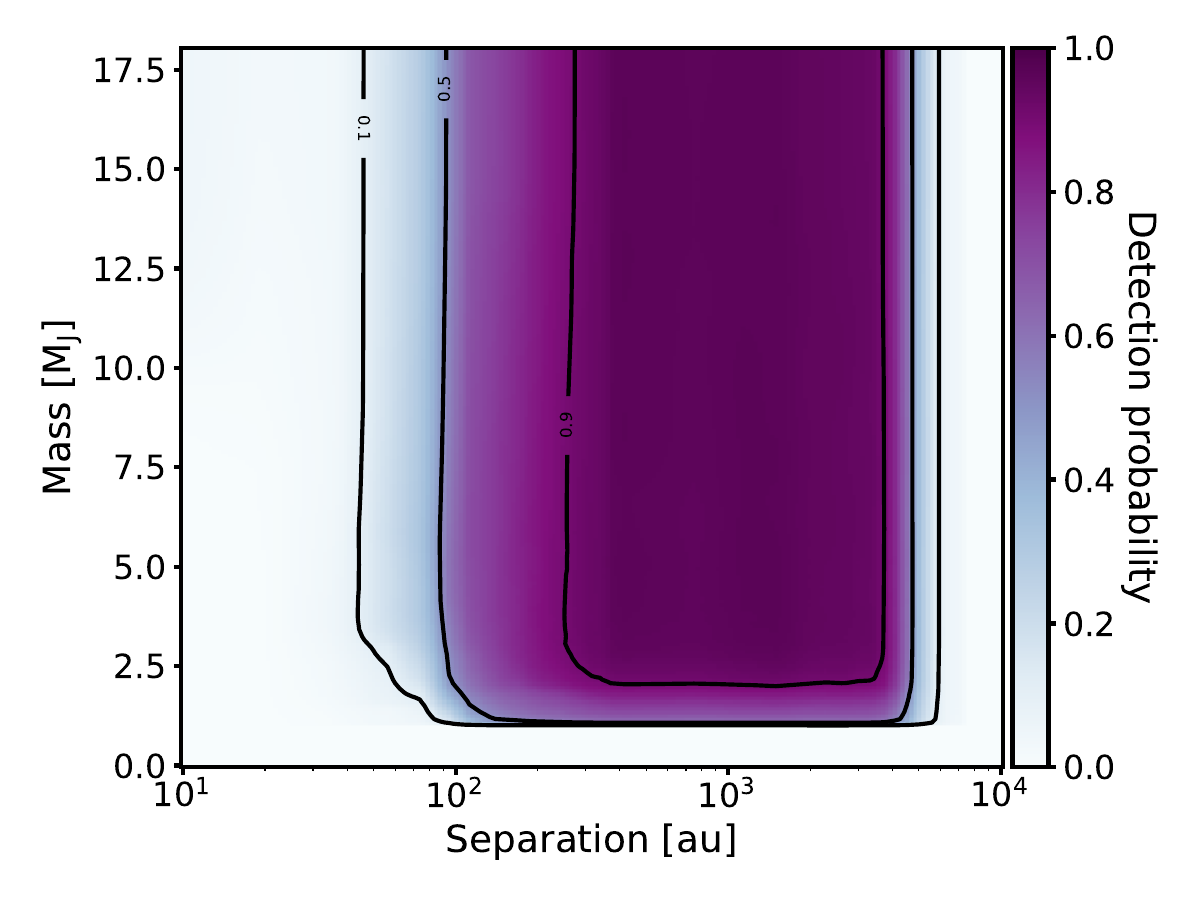}
    \includegraphics[width=1\columnwidth]{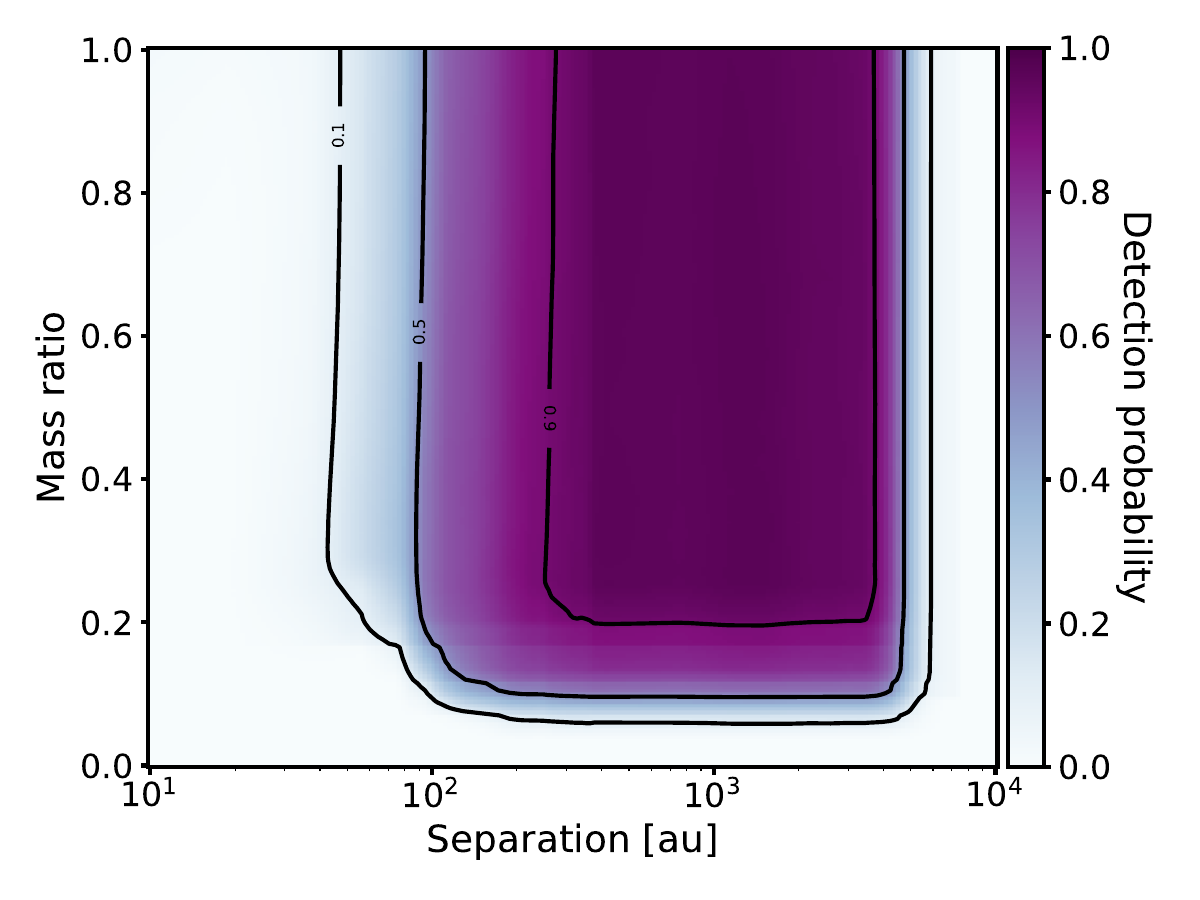}
    \caption[Average completeness maps in different physical units]{Average completeness maps in F127M. The black lines show the 90\%, 50\%, 10\% contours of the detection probability. Top left panel: completeness map in units of apparent magnitude in F127M and angular separation; top right panel: completeness map in units of magnitude contrast and angular separation; lower left panel: completeness map in units of mass and physical projected separation; lower right panel: completeness map in units of mass ratio and physical projected separation.}
    \label{fig:taurus_cmaps_mass}
\end{figure*}

Because of the young ages of our targets, we reached the regime where some of the companion mass should be smaller than 1\,\mj{} with the magnitude contrast we achieved in F127M.
Currently, there are no evolutionary models providing photometry in F127M and F139M for objects less massive than 1\,\mj{} at these ages.
Therefore, we applied a cutoff at 1\,\mj{} in the completeness map.
We show the average completeness maps in different physical units in Fig.~\ref{fig:taurus_cmaps_mass}.
We achieved a detection probability higher than 10\% for companions with masses higher than 1\,\mj{} at angular separations between $\sim$0.3 and $\sim$40 arcsec, corresponding to $\sim$40 and $\sim$6000 au.
We reached a detection limit of 24\,mag and magnitude contrast of 8\,mag in F127M at this detection probability level in our surveys.
Due to the cutoff at 1\,\mj{}, the mass ratio limit we achieved is 0.06 at a detection probability of 10\%. The actual mass ratio limit we can reach should be smaller than this value.
For companions with masses higher than 1\,\mj{} at angular separations between $\sim$1.8 and $\sim$2.2 arcsec ($\sim$250\,au to $\sim$3500\,au), we reached a detection probability higher than 90\%.

\subsection{Binary frequency}
With the completeness map, we can estimate the binary frequency of the sample using Bayesian inference.
Due to the lack of real companion detections, we could only place an upper limit on the binary frequency.
We followed the statistical framework used in \cite{Fontanive2018,Fontanive2023} to calculate the binary frequency of this sample.
It is common to use mass ratio instead of mass when dealing with the multiplicity of stars and sub-stellar objects, because mass ratio is a normalised parameter inherently related to their formation.
It is also a way to mitigate the uncertainties from the mass-age degeneracy in the mass conversion \citep{Fontanive2018}. 
Therefore, we use the mass ratio and projected physical separation in our calculation.

We assume distribution models for companion separation and mass ratio, and generate artificial companions drawn from the models.
Then we apply the completeness correction for the number of predicted companions and compare them with the number of detected companions.
The challenging part of the statistical analysis of this sample is what distributions we should assume for separation and mass ratio. 
Our targets are free-floating young brown dwarfs and planetary-mass objects whose formation mechanisms and multiplicity are not well constrained yet. 
\rev{Owing to the distance of the Taurus star-forming region, our observations are primarily sensitive to companions at wide projected separations. The separation and mass distribution of planetary-mass companions, particularly those with masses as low as 1\,\mj{}, remain poorly constrained in this separation regime. We therefore adopt simple, uniform distributions over the specified parameter ranges as a minimally parameterized baseline model. Given the absence of detections, the data provide limited leverage for constraining additional parameters describing the shapes of the underlying distributions. Introducing a more flexible parameterization would consequently yield parameters that are weakly constrained and strongly dependent on the adopted priors, without being justified by the available data.}

Our survey covers a wide range of separations beyond 30\,au, so we adopted a loguniform distribution for separations between 30 and 7000 au.
In reality, the mass ratio is not an independent parameter of separation and primary mass \citep{Duchene2013}.
It has been found that low-mass objects, such as brown dwarfs, are more likely to have similar-mass companions at small separations \citep[e.g. $<$10\,au;][]{Burgasser2007,Bernat2010,Offner2023}.
But since we are sensitive only to companions beyond 30\,au here, we probably probe the long flat tail of the mass ratio distribution.
We adopted a uniform distribution for the prior of the mass ratio between 0.1 and 1.

Following \cite{Fontanive2018}, we generate a large number of companions ($n$) and calculate the predicted number of companions to be detected in our survey ($k$) by this equation:
\begin{equation}
    k = \sum_{i=1}^{n} (p_i) \times f \times \frac{N_*}{n},
\end{equation}
\noindent where $p_i$ is the detection rate of the generated companion which is calculated in the completeness map, $f$ is the binary frequency, and $N_*$ is the sample size.
The likelihood function is defined by a Poisson distribution:
\begin{equation}
\mathcal{L}(d \mid k) = \frac{k^{d} \, e^{-k}}{d!},
\end{equation}
where $d$ is the observed number of detections and $k$ is the mean expected number of detections from the previous equation.
In the case of a null detection, $d$ = 0.

\begin{figure}
    \includegraphics[width=1\columnwidth]{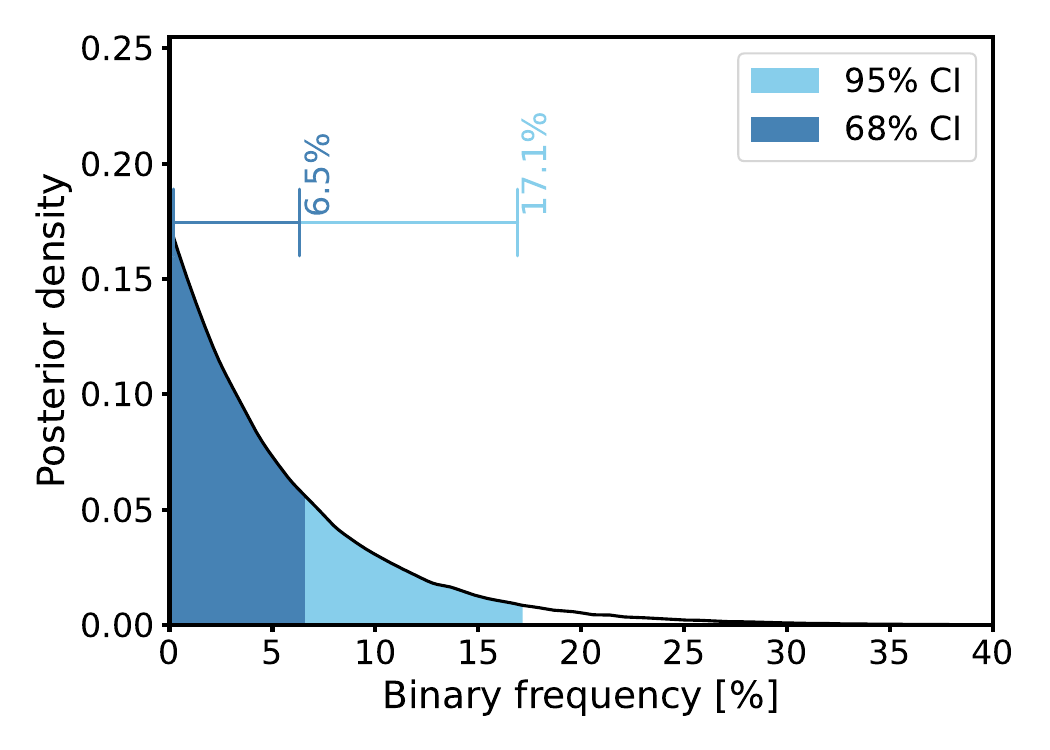}
    \caption[binary frequency of this survey]{Posterior distribution of binary frequency in our null detection survey. `CI' stands for confidence interval.}
    \label{fig:taurus_binary_rate}
\end{figure}

We sampled the likelihood function with the Markov Chain Monte Carlo (MCMC) method, using the \textsc{Python} package \textsc{emcee} \citep{ForemanMackey2013}. We set $n$ = 200,000 and a uniform prior for the binary frequency between 0 and 1.
The posterior of the binary frequency is shown in Fig.~\ref{fig:taurus_binary_rate}.
With null detection, the upper limit of the binary frequency is 6.5\% in the 68\% confidence interval and 17.1\% in the 95\% confidence interval.

\section{Discussion}
\label{sec:taurus_dis}

\subsection{Binary frequency in Taurus and Upper Scorpius}
\begin{figure*}
    \includegraphics[width=1\columnwidth]{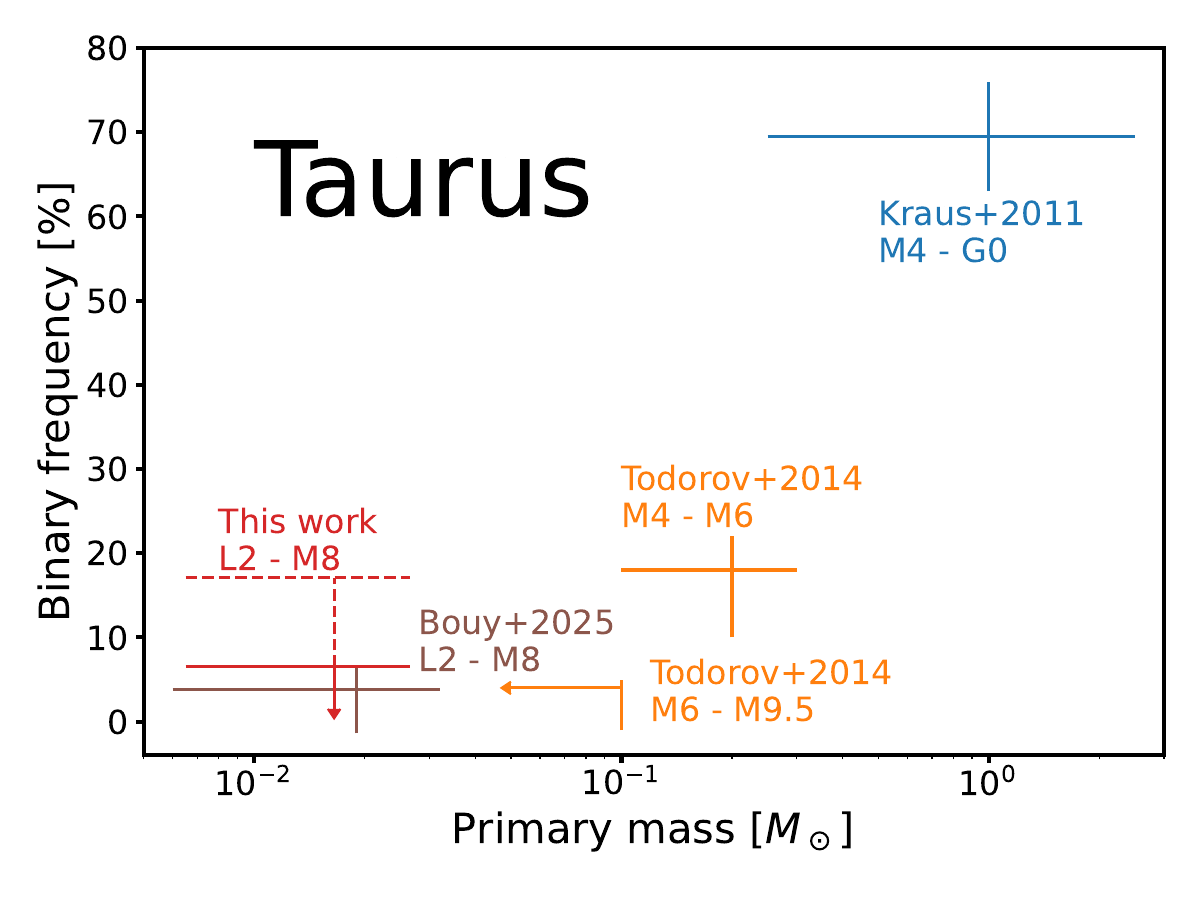}
    \includegraphics[width=1\columnwidth]{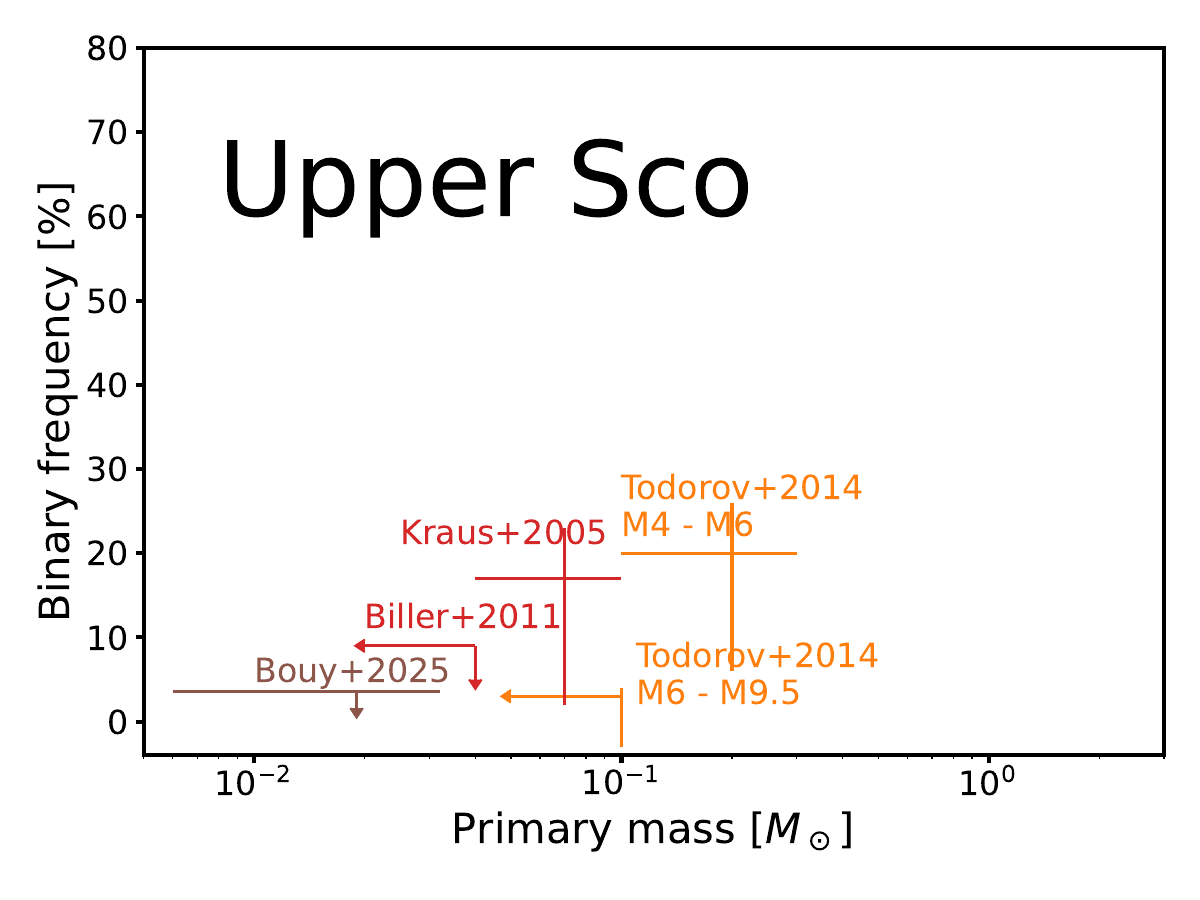}
    \caption[binary frequency in Taurus and Upper Sco]{Left panel: binary frequency as a function of primary mass in Taurus; right panel: binary frequency as a function of primary mass in Upper Scorpius. The binary frequency from the listed literature is not corrected for incompleteness, and their separation resolution limit is between 3 and 10\,au: \cite{Kraus2005,Kraus2011,Todorov2014,Bouy2025, Biller2011}. We used 1$\sigma$ upper limit with a dashed line showing the 2$\sigma$ upper limit when applicable.}
    \label{fig:taurus_upperscor_binary_rate}
\end{figure*}

Figure~\ref{fig:taurus_upperscor_binary_rate} presents the binary frequency measured in Taurus as a function of primary mass.
The binary frequency reported in \cite{Kraus2011} is the fraction of solar-type stars that have at least one companion beyond 3\,au measured with ground-based AO observations.
The binary frequency reported in \cite{Todorov2014} and \cite{Bouy2025} is beyond 10 and 7\,au surveyed with HST imaging, respectively.
While the binary frequency reported in our work is corrected for incompleteness and is calculated for the parameter range of mass ratio in the range of [0.1, 1] and separation between 30 and 7000\,au, the binary frequency from the literature \citep{Kraus2011,Todorov2014,Bouy2025} is not corrected for incompleteness and does not have clearly defined mass ratio and separation range.
These binary frequencies are calculated as the number of detections divided by the total number of targets, which could be slightly higher after being corrected for completeness.
Nevertheless, the binary frequency in Taurus decreases when the primary mass decreases from solar mass to planetary mass ($>$5\,\mj{}), agreeing with the observed trend for field objects \citep[e.g.][]{Offner2023}.
As Taurus is a sparse star-forming region with an age of only 1--10\,Myr for most objects, this suggests that the decreasing binary frequency with primary mass is settled at a very early stage of star formation.
The low binary frequency of brown dwarfs and planetary-mass objects is therefore likely determined by their formation mechanisms.  

Both our work and that of \cite{Bouy2025} provide the first measurements of binary frequency for young planetary-mass objects of similar masses and spectral types in Taurus with HST/WFC3.
\cite{Bouy2025} surveyed 33 objects with masses between 5 and 33\,\mj{} in UVIS filters of F814W and F850LP.
They detected one companion candidate with a mass of 3--6\,\mj{} at a separation of 18\,au and calculated the binary frequency of 3.8$^{+5.2}_{-2.8}$\% including this candidate.
While there are no overlaps between the two surveys, the binary frequency agrees with each other.
The binary frequency of planetary-mass objects is extremely low at wide separations as no companions are detected over 30\,au for the total of 59 targets surveyed in the two surveys.
This indicates that it is intrinsically rare to form planetary-mass binaries, for instance, the direct molecular collapse or filament or disc fragmentation is not efficient and/or the binary systems are extremely vulnerable to dynamical disruptions. 

Upper Scorpius (Upper Sco) is another star-forming region in the solar neighbourhood with an age of $\sim$11\,Myr \citep{Pecaut2012}, and is frequently studied for multiplicity.
We present the binary frequency measured for objects with SpT later than M4 in Upper Sco from \cite{Kraus2005,Biller2011,Todorov2014,Bouy2025} in the right panel of Fig.~\ref{fig:taurus_upperscor_binary_rate}.
Similar to Taurus, Upper Sco also shows a descending trend as the primary mass goes down. 
We observe no significant difference in the binary frequency of brown dwarfs and planetary-mass objects between Taurus and Upper Sco.
Taurus is a low-density star-forming region, and therefore, dynamical interactions between objects are not frequent after their formation. 
Widely-separated low-mass binaries are of low-binding energies and were thought to be likely to be disrupted after 10\,Myr \citep[e.g.][]{Todorov2014}.
\cite{Parker2025} found that a large fraction of planetary-mass binaries would be destroyed within a few Myrs in dense star-forming regions such as the Orion Nebula Cluster.
Since Upper Sco is a denser and older region than Taurus, this suggests that the low binary frequency of these objects should be settled at an earlier stage, e.g. $<$1\,Myr.

\subsection{Binary frequency as a function of age}
\begin{figure*}
    \includegraphics[width=1\columnwidth]{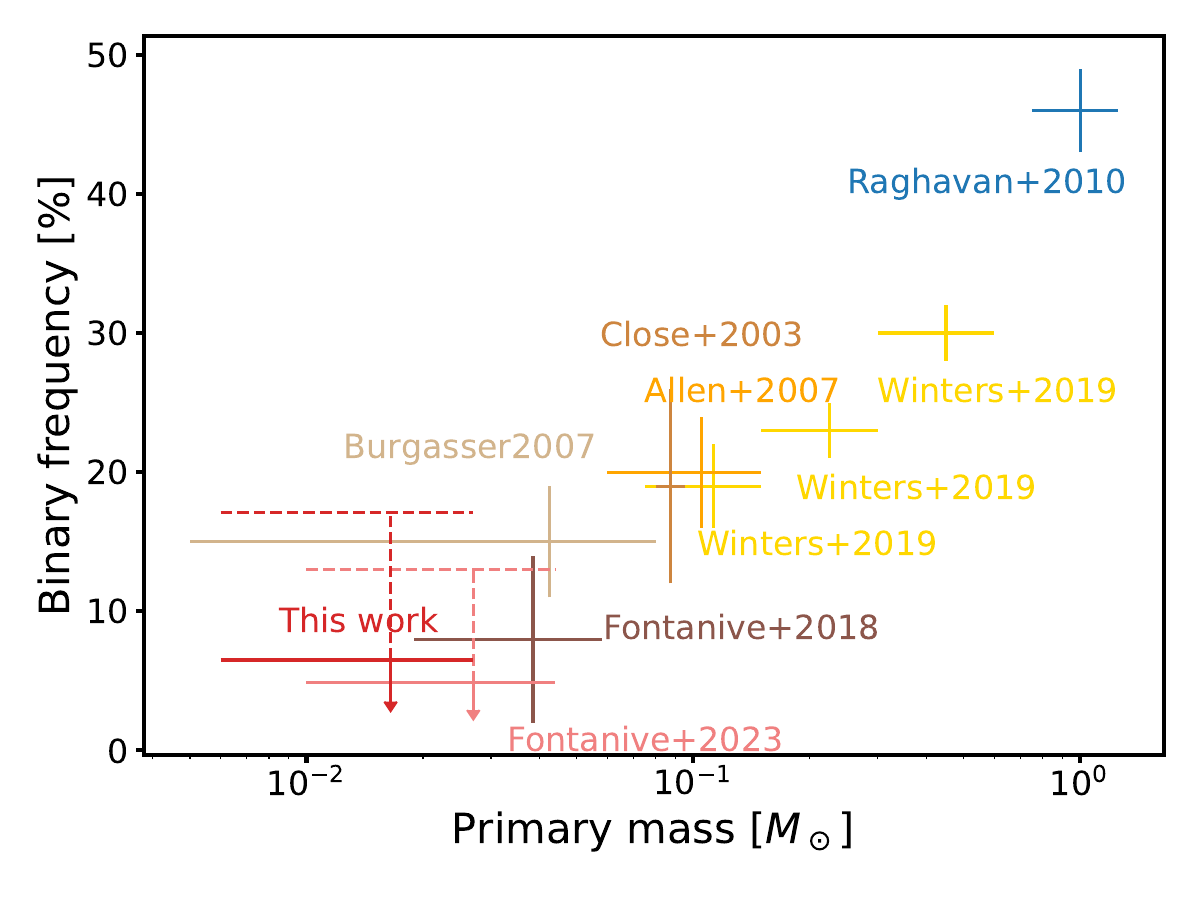}
    \includegraphics[width=1\columnwidth]{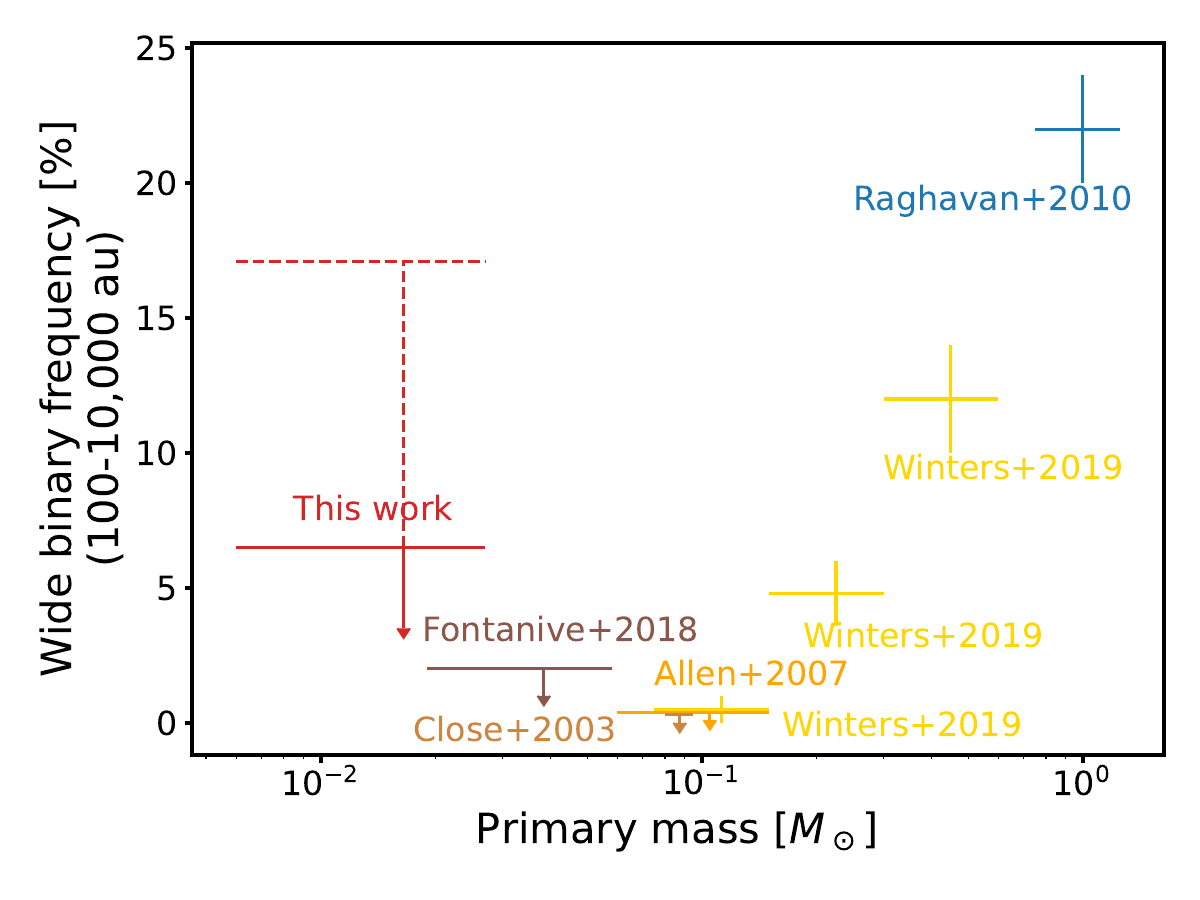}
    \caption[binary frequency as a function of primary mass]{Left panel: overall binary frequency as a function of primary mass for field objects; right panel: wide binary frequency as a function of primary mass for field objects. We also include our results for young planetary-mass objects for the purpose of comparing with field planetary-mass objects.
    The data are taken from \cite{Offner2023} which are corrected for incompleteness. The original surveys are reported in \cite{Close2003, Allen2007,Burgasser2007,Raghavan2010,Winters2019,Fontanive2018,Fontanive2023}. We used 1$\sigma$ upper limit with a dashed line showing the 2$\sigma$ upper limit when applicable.}
    \label{fig:taurus_allbinary_rate}
\end{figure*}

According to evolutionary models, the young planetary-mass objects in our survey will cool to objects with SpT of T8 to Y2+ when they are older than 1\,Gyr.
Comparing the binary frequency of young objects and objects in the field of the same mass can probe the dynamical stability of planetary-mass objects.
We present the binary frequency of field objects as a function of primary mass together with the binary frequency measured in this work in Fig.~\ref{fig:taurus_allbinary_rate}.
We applied the same data reduction and statistical analysis to our sample as \cite{Fontanive2018,Fontanive2023} did to a similar sample of field objects.
The binary frequency of young planetary-mass objects in Taurus does not exhibit a discernible difference compared to the late-T and Y dwarfs in the field surveyed by them.
\cite{Biller2011} also found that the binary fraction of brown dwarfs in Upper Sco is of no significant difference from that of field T dwarfs.
Therefore, with the current small sample statistics and large uncertainty of the binary frequency, we cannot draw a conclusive summary of the trend of binary frequency as a function of age, although both the field and young planetary-mass objects clearly have low binary frequencies.

Since our survey is sensitive to binaries with wide separations such as 2M1207b \citep{chauvin2004, ZhouY2016} and Oph 98AB, we also compared the wide binary fraction at separations between 100 and 10000\,au in the right panel of Fig.~\ref{fig:taurus_allbinary_rate}.
The fraction also drops with the decrease in the masses of field objects.
The same as the overall binary frequency, the difference between young and field brown dwarfs is not evident as only the upper limit of wide binary frequency is constrained.
Actually the separation of the majority of the discovered sub-stellar binaries is within 15\,au, such as WISE J033605.5-014350.4 \citep[a close Y+Y binary][]{Calissendorff2023} and WISE J014656.66+423410.0 \citep[a spectroscopic binary at T/Y transition][]{Bardalez2025}, indicating that the peak of the separation distribution is at tight separations and it is rare to have wide planetary-mass binaries.

The multiplicity surveys of planetary-mass objects are mainly direct-imaging surveys, which are only sensitive to separations larger than $\sim$10\,au for objects in young moving groups and star-forming regions. 
Given that planetary-mass binaries are more likely to form at small separations, spectroscopy and interferometry surveys are necessary to probe tight binaries (e.g.$<$1\,au).
However, these kinds of surveys are subject to the faintness of these objects.
As they are bright in the mid-infrared, the James Webb Space Telescope (JWST) can detect more tight planetary-mass objects, as well as future Extremely Large Telescope (ELT) instruments such as Mid-infrared ELT Imager and Spectrograph \citep[METIS][]{Brandl2021}.
The demography of planetary-mass binaries, to be completed in the future, will reveal the detailed separation and mass ratio distributions of these binaries, as well as the binary frequency as a function of age, which is crucial for understanding the formation of low-mass brown dwarfs and planetary-mass objects.

\section{Conclusion}
\label{sec:taurus_con}
We conducted an HST/WFC3 direct imaging survey of 26 young brown dwarfs and planetary-mass objects with SpT later than M8 and masses between 6 and 37\,\mj{} in the field of Taurus. 
We presented three new young brown dwarfs discovered in the Water Band Infrared Survey and updated the membership of 18 WBIS sources with proper motion measurements.
We performed photometry analysis of the 26 targets in the F850LP, F127M, and F139M filters and estimated their masses.
We constructed the $Q_\mathrm{HST}$ value by combining flux measured in the three filters: objects with SpT later than late-M will have $Q_\mathrm{HST}$ $>$ 1, while background stars should have $Q_\mathrm{HST}$ $<$ 1.
Together with PSF fitting conducted with \textsc{KS2}, no companion candidate is detected at separations over 30\,au.
Artificial star injection tests were performed to build the completeness maps of our survey. 
Using Bayesian statistics, we constrained the upper limit of the binary frequency to 6.5\% in a 68\% confidence interval after completeness correction for mass ratio between 0.1 and 1 and separation between 30 and 7000\,au. 
We confirm that binary frequency decreases smoothly from solar-type stars to the planetary-mass regime in Taurus.
Similar as planetary-mass objects in the field, it is rare to find these low-mass binaries at wide separations. 
No significant difference in binary frequency is detected between planetary-mass objects in Taurus, Upper Sco, and the field, suggesting that the deficiency in planetary-mass binaries of wide separations is determined at a very early stage during formation in a low-density environment.
A larger sample, combined with future surveys of different star-forming regions, will provide more clues as to how the binary frequency varies with age and stellar density, which is essential for understanding their formation.
%
%












\section*{Acknowledgements}

We thank Jonathan Gagn\'e for running Banyan $\Sigma$ using his MOCA database on our sample.
Some of the data presented in this paper were obtained from the Multimission Archive at the Space Telescope Science Institute (MAST). STScI is operated by the Association of Universities for Research in Astronomy, Inc., under NASA contract NAS5-26555. 
This work is based on observations made with the NASA Infrared Telescope Facility, which is operated by the University of Hawaii under contract 80HQTR24DA010 with the National Aeronautics and Space Administration.
Based on observations obtained at the international Gemini Observatory, a program of NSF's NOIRLab, which is managed by the Association of Universities for Research in Astronomy, Inc. (AURA) under a cooperative agreement with the National Science Foundation.
This research made use of \textsc{Astropy}, a community-developed core \textsc{Python} package for Astronomy \citep{Astropy2013,Astropy2018}, \textsc{SciPy} \citep{Scipy2020}, \textsc{NumPy} \citep{Numpy2020} and \textsc{Matplotlib}, a Python library for publication quality graphics \citep{Matplotlib2007}.

\section*{Data Availability}
All observational data used in this work are publicly available on the databases of the corresponding telescopes.



\bibliographystyle{mnras}
\bibliography{main} 








\bsp	
\label{lastpage}
\end{document}